\documentclass{emulateapj}
\shorttitle{The late-type stellar content of the M31 halo}
\shortauthors{A. Koch \& R.M. Rich}
\slugcomment{Accepted for publication in the Astronomical Journal}
\def\col{{\em 81}$-${\em 77}}

\begin{document}

\title{A statistical analysis of the late-type stellar content in the Andromeda halo}

\author{Andreas Koch\altaffilmark{1}, and R.~Michael Rich\altaffilmark{2}}
\email{ak326@astro.le.ac.uk; rmr@astro.ucla.edu}
\altaffiltext{1}{Department of Physics and Astronomy, University of Leicester, University Road, 
Leicester LE1 7RH, UK}
\altaffiltext{2}{UCLA, Department of Physics and Astronomy, Los Angeles, CA, USA}
\begin{abstract}
We present a statistical characterization of the carbon-star to M-giant (C/M) ratio in the halo of M31. 
Based on application of pseudo-filter band passes to our Keck/DEIMOS spectra we 
measure the \col-color index of 1288 stars in the giant stellar stream and in halo fields out to large distances.  
From this well-established narrow-band system, supplemented by V$-$I colors, we find only a low number (five in total) of C-star candidates. 
The resulting low C/M ratio of 10\% is consistent  with the values in the M31 disk and inner halo from the literature.  
Although our analysis is challenged by small number statistics and our sample selection, there is an indication that 
the oxygen-rich M-giants occur in similar number throughout the entire halo. We also find no difference in the C-star population of the 
halo fields compared to the giant stream. 
The very low C/M ratio is at odds with the observed low metallicities and the presence of intermediate-age stars at large radii. 
Our observed absence of a substantial carbon star population in the these regions indicates that the (outer) M31 halo cannot be dominated by the 
debris of disk-like or SMC-type galaxies, but rather resemble the dwarf elliptical NGC~147. 
\end{abstract}
\keywords{stars: carbon --- stars: late-type --- Galaxies: evolution --- Galaxies: stellar content --- Galaxies: structure --- Galaxies: individual (\objectname{M31})}
\section{Introduction}
As more observations accumulate, it has become increasingly clear that the field population of massive galaxy halos
 consists largely of the debris of accreted satellites, along the lines of  Searle \& Zinn (1978).  In the case of the
Milky Way, the orbital stream of the Sagittarius (Sgr) dwarf spheroidal (dSph) is so complicated that it is easy to imagine its members
being dispersed throughout much of the halo over a few Gyr timescale (e.g., Law et al. 2005; Fellhauer et al. 2006).  Additional observations find coherent structures
and satellites over much of the Galactic halo (e.g., Belokurov 2007a,b).    Evidence of  mergers  in the form of the giant stream and other structures, is clearly even more striking in M31 (Ibata et al. 2001, 2007; Ferguson et al. 2002).  N-body models of these events show that even a single minor merger is capable of filling much of the halo with debris after only 2 Gyr (Mori \& Rich 2008).   One is also struck by both the surface
brightness and spatial extent of the giant stream and its shell feature, all of which are attributable to one relatively recent minor merger, that did not disrupt the disk (Fardal et al. 2006, 2008; Mori \& Rich 2008).    The importance of the giant stream event emphasizes that, 
as halo histories are stochastic, the populations of halos may be dominated by the debris of just a few events involving massive satellites -- certainly not the picture that one may infer from the study of their globular cluster systems.

In the case of M31, direct measurements of the halo star formation history from the main sequence turnoff revealed a complex
range of age and abundance (e.g., Brown et al. 2003, 2006, 2008).  In all of the M31 halo fields, there exist significant
populations of intermediate-age stars, with metallicities ranging from 1/10 to twice Solar.   In addition to a
contingent of stars ranging from 5--10 Gyr, nearly all M31 halo fields show signs of a blue
extension of $\sim 1$ Gyr old stars (e.g. Brown et al. 2006; Richardson et al. 2008).   If stellar systems are being ingested 
then  their properties should, in principle, reflect those of similar galaxies in the Local Group.   Although star formation likely ceases
as a satellite is tidally disrupted, it is clear that much of the complexity can be observationally reconstructed, as has been
accomplished for the Sagittarius dSph galaxy (e.g., Chou et al. 2007).

An efficient tracer of more recent accretion events through intermediate-age populations is the number ratio, C/M, of carbon-rich (C) stars to the oxygen-rich (M) stars.  
C- and M-type stars are the most luminous single stars found in intermediate-age populations, and their spectral signatures are distinctive.  In complex stellar systems, the presence of carbon stars in significant numbers guarantees the presence of substantial intermediate-age populations. 
 The primary correlation in  the C/M ratio is with metallicity (Iben \& Renzini 1983; Brewer et al. 1995); in more metal rich systems, the balance tilts toward the oxygen-rich M-stars.   For example, the Galactic bulge is dominated by M giants (Blanco et al. 1984).   Intermediate-age populations are carbon star dominated, with this tendency to increase at lower metallicity (e.g. Groenewegen 2002; Mouhcine \& Lan\c con 2003; Batinelli \& Demers 2005).   
 Likewise, radial age and/or metallicity gradients in a given stellar system have been confirmed to be accompanied by  gradients in the C/M ratio (e.g., 
 Cioni et al. 2008). 
 Furthermore, the Small Magellanic Cloud is well known to contrast with the LMC in having a high C/M ratio (Blanco et al. 1984).   Although carbon stars are known in star clusters of the Magellanic Clouds (see e.g. Mould \& Aaronson 1979), no late-type, luminous carbon stars are found in Galactic globular clusters despite there being a large contingent of massive, metal poor clusters.   
 It may be inferred that the deep mixing on the AGB necessary to produce a carbon star is only possible for intermediate age stars. 
 
 Most studies of the C/M ratio in the Local Group rely on imaging through narrow band filters that measure the relative strength of CN and TiO bands (e.g., Cook et al. 1986).  In this context, the \col~-color index is a measure for the relative strengths of CN and TiO bands and thus efficiently segregates 
M-stars with strong TiO and C-stars with strong CN bands. Quantitative analyses require additional color information (such as V$-$I), 
in order to construct distinctive two-color diagrams (TCDs). 
Specifically, C-Stars in M31 have been recorded (Brewer et al. 1995; Battinelli et al. 2003; Demers \& Battinelli 2005; Battinelli \& Demers 2005). The inferred relatively low C/M ratios 
of the order of 0.1 in the M31 disk and the inner halo (within 10 kpc) are then in agreement with the metal rich character of these populations. 

For  the present work, we have examined spectra from a large survey of the M31 halo (Koch et al. 2008; hereafter K08), and we have synthesized the CN and TiO filters for our spectra.   From those we explore the C/M ratio in M31 halo fields, some of which are further than 100 kpc from the nucleus, but we do so without the completeness of a magnitude-limited photometric survey.  Even considering these limitations, we can statistically investigate the late-type stellar content in the Andromeda halo and connect occurrences of either population to the prevailing ages and metallicities out to large radii. 
This Paper is organized as follows: 
In \textsection 2 we briefly summarize our observations and the standard reductions, while 
\textsection 3 explains our definitions of the CN$-$TiO pseudo-filters. 
The  separation into C- and M-stars from these filters, possible biases, and the implications for M31's halo populations are  
discussed in \textsection 4.  Finally, \textsection 5  summarizes our findings.  

\section{Observations and reduction}
The data used in this work were collected 
in the course of a  Keck program focusing on the spectroscopic investigation of  M31's 
halo structures based on the kinematics and chemical analysis of red giants. 
Since these data are identical to those presented in K08, we refer the reader to 
the latter (and references therein) for details on the target selection, data collection and reduction 
for the whole project.

To recapitulate, our data set covers fields on the south-east minor axis of M31 at projected distances of 
9 kpc out to 160 kpc\footnote{Adopting a distance to M31 of 784 kpc (Stanek \& Garnavich 1998).}.  
These fields were observed using the DEIMOS multislit spectrograph at the 
Keck\,II 10\,m telescope over several observing runs from 2002 through 2006.  
The majority of the spectra was centered at a wavelength of 7800\AA, yielding  
a full spectral coverage of $\sim$6500--9200\AA, which also comprises the prominent 
molecular TiO and CN bands that are of interest for the present study. 

The photometry of our targets comes from two sources: 
For the outer halo fields (R$\ga$25 kpc) the Washington $M$, $DDO51$ and $T_2$ photometry of 
Ostheimer  (2003) was transformed into standard Johnson-Cousins 
V and I magnitudes following Majewski (2000). 
Stars from this photometry were pre-selected to maximize the number of red giants, combining their location in the color magnitude diagrams (CMDs) and TCDs built from the Washington filters so as to optimally avoid the dwarf sequence  (i.e., blueward of the red giant branch [RGB],  and {\em much} brighter than the tip of the RGB [TRGB]; e.g., Majewski et al. 2000; Palma et al. 2003), and to cover the entire RGB so as to yield a broad, unbiased metallicity range.

The photometry of the inner M31 fields  (R$\la$25 kpc)  was, on the other 
hand, taken from the MegaCam/MegaPrime archive (Gwyn 2008) 
and transformed to the Johnson-Cousins magnitudes via the latest  Padova stellar 
isochrones (Marigo et al. 2008),  which are available both in the CHFT photometric system 
and  for Johnson-Cousins V and I. 
As above, the goal of the respective target selection was to avoid those  regions of the CMD with the greatest foreground dwarf contamination (and to exclude galaxies based on morphological criteria; e.g., Gilbert et al. 2006). With the aim of targeting bright stars to reach sufficient S/N ratios, highest priority was given to stars with $20.5\la I_0\la 22$ mag. 
Since M31's TRGB lies at approximately 20.5 mag (e.g., Durrell et al. 2001), some bright AGB stars above the TRGB will have been removed from the initial samples.  
In fact, our sample retains a number of bright (red) stars above the RGB tip, so we conclude that our relative completeness does not fall so rapidly. The redder stars are actually fainter in the I-band due to molecular absorption, thus permitting bolometrically bright stars to fall within the I-band selection (see also CMDs in Fig.~5). 
Targets selected from either survey reach similar limiting magnitudes (slightly fainter than  22.5 mag) and span a comparably broad color range, so we conclude that no selection criterion has a bias set either in selecting C- or M-stars. 

Based on stellar radial velocities, the equivalent widths of the gravity-sensitive Na doublet
and the stars' V$-$I colors, we  obtained an efficient separation of M31 red giants from 
the Milky Way foreground contamination in K08.  
K08 reject 781 stars as being likely foreground dwarfs, while 1288 were classified as likely M31 giants and comprise our present sample.  
\section{Spectroscopic filter definition}
The narrow band ($\Delta\lambda\sim$300\AA)   {\em 77} and {\em 81} filters 
centered on molecular bands have proven successful in separating C and M 
stars in Local Group galaxies, when coupled with temperature sensitive broad-band photometry. 
In particular, the red giant branch (RGB) bifurcates into the redder (in \col) C-stars and the 
M-stars with V$-$I redder than $\sim$1.5--2.0 (Cook et al. 1986; Cook \& Aaronson 1989; Nowotny 2001, 2003; Harbeck et al. 2004) with similar cuts for other color indices 
(e.g., R$-$I; Battinelli et al. 2003; Demers \& Battinelli 2005). 

Our program for K08 was originally designed to focus on spectroscopy so that no photometry in selected narrow-band  
filters was taken. Thus we have to rely on integrating our spectra in band passes mimicking the traditional filter curves. 
Depending on the choice of telescope and instrument, the exact filter curves will differ and therefore yield slightly different zero points and color indices. 
We investigated the different instrumental set ups by convolving our spectra with filter curves from those three sites that are most widely found in the literature on the late-type stellar content of Local Group galaxies: Firstly, the CFHT filters\footnote{\url{http://ftp.cfht.hawaii.edu/Instruments/Filters/megaprime.html}} 
as used by, e.g., Battinelli et al. (2003); Demers \& Battinelli (2005), to name a few; 
secondly, the band passes available for the WIYN telescope\footnote{\url{http://www.wiyn.org/new2006/observe/filters.html}} (e.g., Harbeck et al. 2004); and 
finally,  we applied  Gaussian filter curves (Cook \& Aaronson 1989) 
as defined in the Asiago data base (Moro \& Munari 2000; Fiorucci \& Munari 2003). 
The respective transmission curves are compared in Fig.~1.
\begin{figure}[tb]
\begin{center}
\includegraphics[angle=0,width=1\hsize]{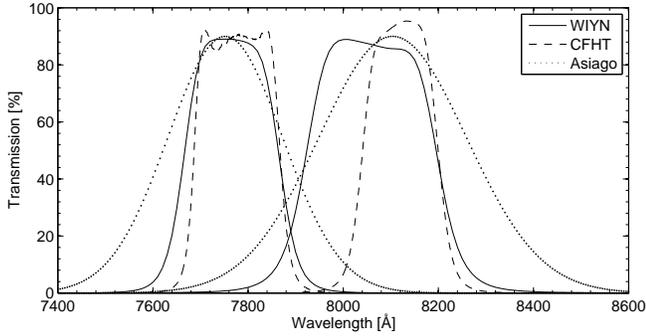}
\end{center}
\caption{Transmission curves of the narrow-band TiO (at 7750\AA) and CN (at 8100\AA) filters, from various telescopes and data bases.}
\end{figure}

While the {\em 77}-filter curve, centered at 7750\AA, measures the TiO band head in M-stars, it essentially 
samples the continuum in C-star spectra. The {\em 81}-filter at 8100\AA, 
on the other hand, is a measure for CN absorption in the C-stars and covers the stellar continuum 
of the cool M-giants. This is illustrated in 
Fig.~2 for two stars from our M31 sample that were classified as C- and M-stars based on the 
\col-color, and for a background object with essentially flat continuum in either filter. 
\begin{figure}[tb]
\begin{center}
\includegraphics[angle=0,width=1\hsize]{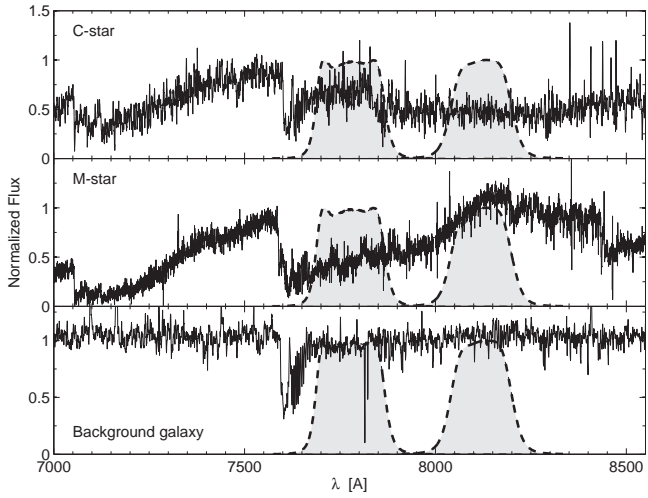}
\end{center}
\caption{Sample spectra of  a C-star candidate (top panel), a cool M-giant (middle) and 
a background galaxy (bottom). The latter, absorption-free spectrum illustrates the  
overlap with the atmospheric $A$-band that contributes the only absorption within 
the band passes. 
Both stars have radial velocities consistent with membership with M31. 
The spectra were arbitrarily normalized at 7562\AA. 
Indicated as dashed lines are the CFHT-based filter curves we used for measuring the 
\col~color index.}
\end{figure}
The low signal-to-noise (S/N) ratios of these examples already indicates that a purely 
spectroscopic classification of our sample would be difficult to achieve; therefore the main objective of this   
pseudo-photometric study is rather a statistical endeavor (note that the median S/N of the entire sample 
in K08 is a mere 8.5 per pixel). 

In practice, we define 
\begin{equation}
{\it 81-77} = -2.5\,\log\left( \frac{\int F(\lambda)\,G_{\it 81}(\lambda)\,d\lambda }{
                                                       \int F(\lambda)\,G_{\it77}(\lambda)\,d\lambda } \right) \, + \, const., 
\end{equation} 
where $G_i$ denotes the filter profiles that were taken from the aforementioned telescopes' filter characteristics. 
The integration is carried out over the whole Doppler corrected spectral flux, $F$. 
The constant in eq.~1 accounts for the integral over 
the transmission curves and amounts to  ($-$0.086, 0.351, 0.272) for the (CFHT, WIYN, Asiago) filtes, respectively. We note, however,  that an accurate  
zero point is not required for this study, since we are primarily interested in an empirical identification of the different stellar populations. 

Errors on the color were determined by Monte Carlo variations of the spectra accounting for 
the spectral noise and uncertainties in the radial velocities (entering the measurements upon 
Doppler shifting the spectra). While the velocity contribution is negligible, at a median 
$\sigma_{\it 81-77}$ of 0.0001 mag, the random error due to the S/N dominates the error budget 
and yields a  median total random color uncertainty of 0.012 mag (with 95.7\% of the 
errors smaller than 0.04 mag).  
It turns out that all three different filter definitions yield very similar results and the resulting color indices are extremely well correlated. 
Fig.~3 shows a comparison of each narrow band measurement and despite the fact that the trends are not strictly unity, there is only 
little scatter of less than 0.1 mag (1$\sigma$) around the linear relations. 
\begin{figure}[tb]
\begin{center}
\includegraphics[angle=0,width=1\hsize]{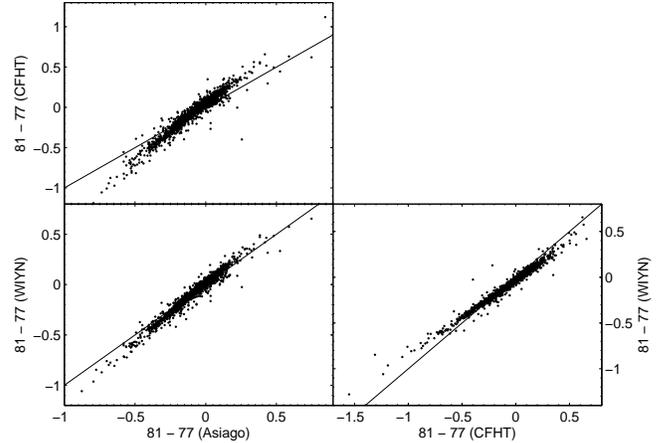}
\end{center}
\caption{Comparison of the narrow band \col-index obtained from integration of the spectra using three different filter definitions.}
\end{figure}
However, a major source of uncertainty is the presence of telluric absorption that falls into the  
filter band passes (Fig.~2, bottom panel). In particular, the {\em 77}-band slightly overlaps the atmospheric $A$-band at 7600\AA,  
which will remove flux from the stellar spectrum that matters for the filter integration (e.g., Cook \& Aaronson 1989). 
This prompted us to adopt the CFHT-based filter definitions for our final analysis, as this particular system has the smallest bandwidth and thus shows 
the least overlap with the A-band (Fig.~2).  

Secondly, the gap between the two adjacent DEIMOS CCDs 
can introduce flux offsets that may affect a measure of the true color index.  
To estimate the bias introduced by these effects, we performed the identical 
integration over the filter curves for a subsample of background galaxies in the M31 sample 
(see K08) with preferentially flat continua.  
To first order, these spectra contain only a flat continuum over the entire spectral range, plus the atmospheric 
contaminants (emission lines that are red shifted into the filter band passes result in a strongly aberrant   
\col~and can be easily discarded).  
As a result, the \col~of the galaxies is compatible with zero as expected, with a 1$\sigma$ scatter 
of 0.10 mag.  An additional test of this error component consists of estimating the scatter in the 
pseudo-color of dwarf stars around $V-I\sim0$ (see also Fig.~4; bottom panel), 
which are expected to contain only the telluric features 
in the spectral region of interest. In concordance with the galaxies, the respective scatter of \col~in the bluer dwarfs amounts to 0.10 as well. Thus we add this value  in quadrature to  the color error to account 
for undesirable telluric absorptions.

\section{The C-M star ratio}
In Fig.~4 we show the TCD of our stars in the sense of CN$-$TiO (\col) versus V$-I$, both for 
giant and dwarf stars according to the separation in K08. In order to select possible C-star 
candidates and M-giants, we draw 
empirical lines that frame these populations (cf. Cook et al. 1986; Harbeck et al. 2004, 
and references therein),  but note that their exact choice remains 
somewhat arbitrary due to potential zero point uncertainties in the colors 
and our non-standard, i.e., spectroscopic derivation of the \col~color.   
M-stars were selected from those diagrams by imposing a cut at V$-$I=2, which strictly 
selects spectral types later than M5 (``M5+''; e.g., Demers \& Battinelli 2003). 

To select C-star candidates, we used a bluer cut at V$-$I of 1.55. Although this means that we do not 
obtain a homogeneous selection of C- and M-stars in the same color space, this cut-off is justified by the 
spectral type assignments of Nowotny et al. (2001), according to which the C-star branch extends relatively bluewards (as indicated by the square symbols in Fig.~4). Moreover,  the presence of stars in this selection box, which are clearly detached from the regular RGB and that are potentially identifiable through their spectra,  
 is in agreement with our bluer C-star candidate criterion.  
The limits
on the \col~color, on the other hand are set to $\ge0.25$ (C) and $\le0$ (M), in accordance with 
the boundaries used by Cook et al. (1986); Battinelli et al. (2003); see also Brewer et al. (1996).
\begin{figure}
\begin{center}
\includegraphics[angle=0,width=1\hsize]{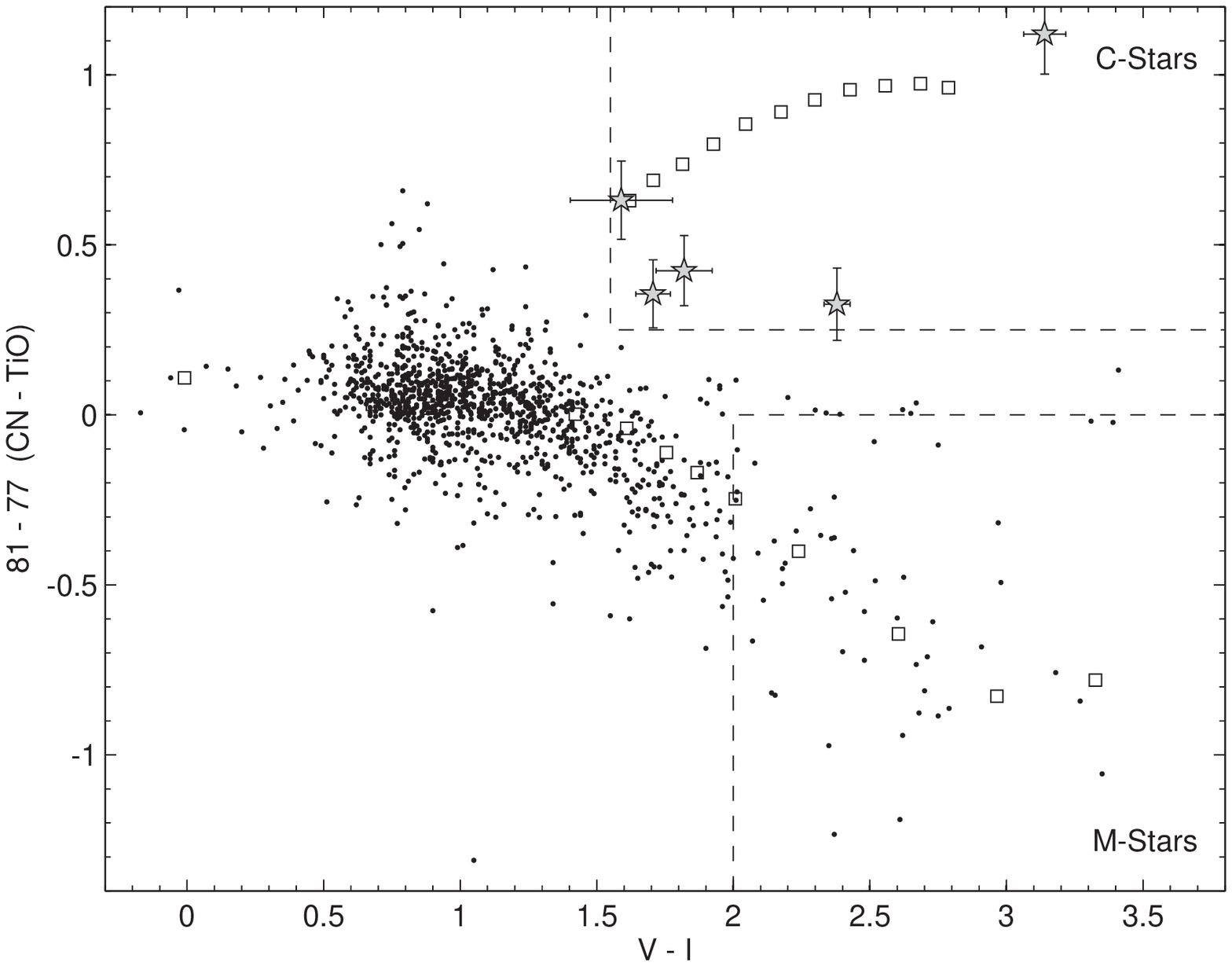}
\includegraphics[angle=0,width=1\hsize]{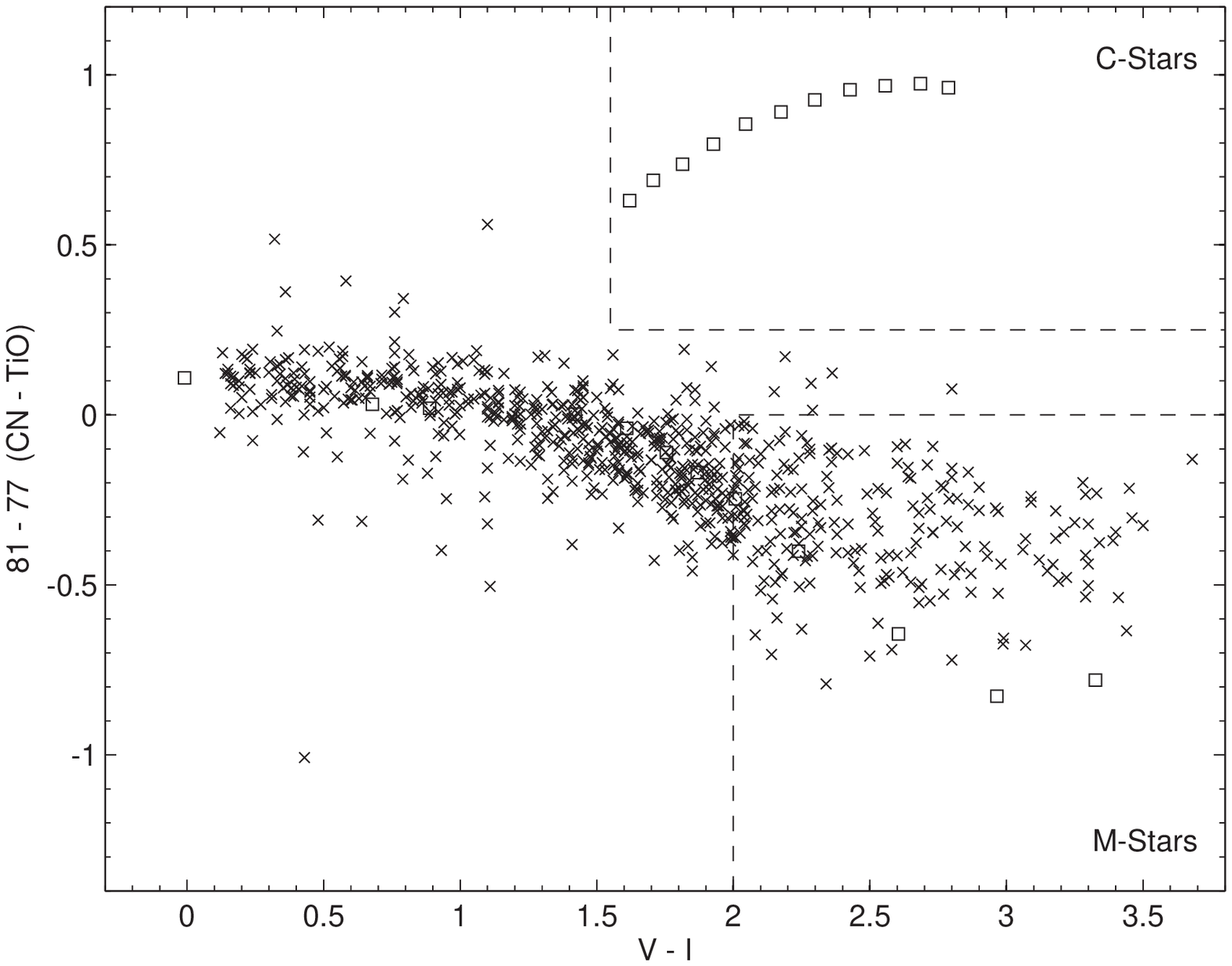}
\end{center}
\caption{Two-color diagram, separated into dwarfs (bottom panel) and giants (top panel). 
Indicated as dashed lines are the empirical sections to separate C- and M-star candidates. Open squares indicate the empirical populations for 
the RGB and carbon stars from Nowotny et al. (2001). The 5 carbon star candidates are highlighted as filled star symbols in the upper panel.}
\end{figure}

Overall, those five C-Star candidates in the selection box are affected by low S/N and suffer from 
unsecure spectral classification. Only for very few cases, smoothing of the spectra yields a weak 
indication  of an excess in the CN band (cf. Fig.~2).
M-stars, on the other hand, are usually bright 
and clearly distinguished by their prominent TiO bands, already upon visual inspection. 
We also show, in Fig.~5 (left panel), the location of the five tentative C-stars in a CMD. Two of those are fainter than I$>$22 mag, while the remainder are comparably bright objects. None of remaining three candidates lies above the TRGB\footnote{We adopt here  an I$_{\rm TRGB}$ of 20.52$\pm$0.05 from Durrell's et al. (2001) study of a halo field at 20 kpc. We note, however, that Ibata et al. (2007; their Fig.~45) estimate a spread in distance modulus of up to one magnitude due to the large depth extent of the M31 halo, which may affect conclusions drawn from stars covering a range of distances out to large radii.}, although the two brightest ones are comparable to  its magnitude. 
Likewise, we show in the right panel of Fig.~5 the 50 M-giant candidates. It is reassuring that they populate the same locus in the CMD as the C-star candidates. 
Since we indeed  detected copious M giants and therefore likely AGB stars in this boundary, this shows that this region is not biased against C stars:  
if present in larger number, we should have been able to detect them regardless of our preselection criteria (Sect.~2).
\begin{figure}
\begin{center}
\includegraphics[angle=0,width=1\hsize]{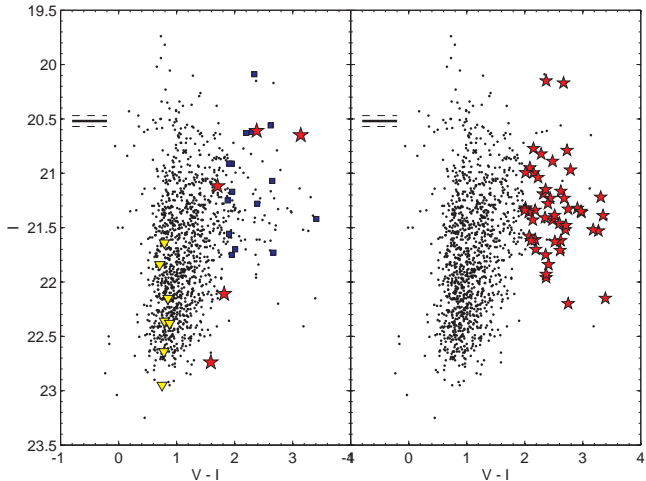}
\end{center}
\caption{CMD of all stars identified as giants in K08 (small dots; see Fig.~3 in K08). The five nominal C-star candidates (Fig.~4, top panel) are highlighted as filled star symbols in the left panel, while the right panel shows stars from the M-star selection box.. 
Horizontal lines indicate the TRGB of 20.52$\pm$0.05 from Durrell et al. (2001). Small filled squares show the ``TCD outliers'' with red V$-$I colors at (CN$-$TiO)$\sim$0, while solid triangles are those stars with large positive (CN$-$TiO) on the blue side of the TCD.}
\end{figure}
\subsection{Andromeda's C/M5+ ratio}
Overall, there are very few C-star candidates found in the respective region of the giants' 
TCD (top panel of Fig.~4). By simply counting numbers in either box, we find a C/M5+ ratio 
for the entire M31 halo sample of (10$\pm$5)\%, with an error  based on Poisson statistics. 
In total, there are 5 (50) objects with colors that {\em formally} imply that they are C- \mbox{(M5+)} stars. 
An accurate account of the measurement errors prompted us to also run a number 
of  Monte Carlo simulations, varying the sample by its uncertainties in either color.  
Thereby we find a similar, statistical value of the C/M5+ ratio of (0.10$\pm$0.03), where 5 (53) stars on average 
are scattered into the C- (M-) star selection region. 
The late-type stars thus make up a fraction of (4$\pm$2)\% 
of our entire spectroscopic red giant sample.

Our measured C/M5+ ratio compares to a foreground-corrected value of 8.4\% found in the {\em outer disk} of M31 
(Battinelli  et al. 2003) and those of  $\sim$10\% by  Demers \& Battinelli (2005) and 
 2\% to 10\%  out  to 35 kpc found by Brewer et al. (1995). Note that all these latter studies   
identify their C-star candidates from narrow band photometry only  and use slightly different CMD selection criteria (see also Sect.~4.3.3). 

\subsection{TCD Outliers}
Two regions in the TCD (Fig.~4) contain stars that 
do not seem  to be compatible with the colors of any population within, nor outside of the selection boxes.
The first group comprises approximately seven stars with large CN absorption compatible with that of C-stars 
(\col~$\sim$0.6 mag), but with V$-$I of $\sim$1 and therefore typical of the RGB.   These stars cover 1.5 mag in magnitude towards the fainter end of the observed RGB (Fig.~5). 
Thus their respective spectra are largely governed by strong noise and, in two cases,  flux offsets between the DEIMOS CCDs so that they were easily discarded. 
It is interesting, though, that Battinelli \& Demers (2005)  find comparable values of positive (CN$-$TiO) on the blue side of their TCD from their photometry at large radii ($>30$ kpc) in the M31 disk.

Secondly, we note a plume of the order of 12 red (V$-$I$\ga2$) stars with a \col~$\sim0.1$, 
which thus scatter away from the well defined RGB. Most of those spectra withstand a classification as late K-type to early M-giants (by visual inspection), thus indicating highly erroneous 
V$-$I,  although their nominal color uncertainties are not unusually large. The remainder of those spectra are largely dominated by noise. 
A possible reason for larger uncertainties in some stars' colors may be their variability on the AGB. As the luminosity amplitude of such stars is much larger in V than in the I-band (as high as 1 mag), their colors can be falsified (in the sense of a broader color scatter)
if the the V and I photometry  were taken at different epochs. 
 Moreover, even if the V and I photometry are obtained nearly simultaneously, as  for this work (and K08), 
V$-$I could be abnormally red if the observations were taken at minimum light. 
Therefore, systematic variability of the reddest of these particular stars cannot be ruled out as an additional error source (see also Battinelli \& Demers 2005). 

All in all, these two components fall outside the ranges populated by the 
standard M- and  C-star populations considered here and they
are unlikely to affect the C/M ratio in the present analysis, given the empirical  
treatment of this ratio and the (photometric) errors.  
\subsection{Potential biasses in C/M}
An accurate determination of the C/M star ratio from a spectroscopic survey as presented here can suffer from several effects: 
\subsubsection{Foreground contamination}
First, a potential residual  contamination with foreground dwarf stars can artificially decrease the C/M 
ratio. As addressed in K08, this component is expected to be largely reduced through a laborious 
treatment of V$-$I-color, velocity and Na doublet equivalent widths as membership criteria. 
To this end, we separated the TCDs in Fig.~4 into foreground dwarfs and M31 giant candidates 
following the assignment of K08.  

While there are no interfering Milky Way dwarf carbon stars to be expected (Totten et al. 2000), as is 
in fact seen in our data (bottom panel of Fig.~4),  
there can be a non-negligible dwarf contamination in the M-star regime. 
On the other hand, this issue is efficiently resolved by just the addition of the dwarf/giant information 
from K08 and we expect no significant residual dwarf component in the ``giant'' sample that 
would falsify the final C/M ratio. Furthermore, Brewer et al. (1996) found that all stars in their TCD selection 
region were spectroscopically confirmed C-stars with no other contaminating spectral types. 
We also note that the sequence of foreground M-giants (bottom panel of Fig.~4) is in full agreement with the 
observed Galactic dwarf component (e.g., Letarte et al. 2002) and the empirical colors of Nowotny et al. (2001), 
which reinforces the accuracy of our separation procedure in K08. 
\subsubsection{Photometric homogeneity}
 In order to investigate to what extent possible inhomogeneities in the input photometry and the mixture of several observing runs, extending 
over a period of $\sim 4$ years (K08) could affect our conclusions, we separated our sample into individual TCDs for each observing run, in particular 
those for the inner fields -- for these, the photometry is based on the homogeneous MegaCam survey only. These diagrams are shown in Fig.~9 in the appendix.
In either case, the expected tracks for the RGB and the M-giant populations are well reproduced and the distribution for the foreground dwarfs is homogeneous between 
the runs as well. Furthermore, the values for the C/M ratio we find from data in different fields are consistent, albeit hampered by overall low numbers (see also Sect.~4.1).  
This reassures us that our use of data from different epochs and observing runs does not incur any inhomogeneities in the evaluation of our C/M ratios. 
Furthermore, when plotting TCDs separately for all fields from MegaCam and all Washington-selected ones, we find no difference (Fig.~10 in the appendix). 
Finally, we resolve the issue of whether potential differences in the limiting magnitudes of both sources for the target (pre-) selection could bias the detectability of C- and M-stars with respect to numbers or their radial distribution. 
Therefore, we construct TCDs and the C/M ratio by imposing several magnitude cuts -- for all samples that obey I$<$21.5, 22.0, 22.5 mag, respectively, the general trends in the diagrams are broadly preserved and the same order of magnitude of $\sim$10\% in the C/M ratio is recovered. We conclude that non-homogeneities between our input sources are negligible and of no concern for the present study.
\subsubsection{Incompleteness}
Finally, the targets for the spectroscopic study were deliberately pre-selected from Washington photometry 
(Ostheimer 2003; Palma et al. 2003) and the MegaCam data to follow the locus of the RGB and to cover a broad part of the giant branch to be unbiassed with regard to metallicity, thereby including some red stars. 
The main goal of K08 was to maximize the number of red giants while avoiding regions of the CMD with the greatest foreground dwarf contamination (i.e., blueward of the RGB,  and much much brighter than the TRGB).  
This means that also the contribution of cool, late-type stars to the present sample is potentially  
incomplete. Strictly, the C/M ratio should be determined using C- and M- AGB stars only, with a faint limit for the AGB M-stars at  M$_{\rm bol} < -3.5$ 
(Brewer et al. 1995), corresponding to the TRGB of M stars. Therefore, the low value of C/M we find would be caused by an overestimate of the number of M5+ AGB stars in our sample. Although we do not have any means to directly uncover this number of ``true'' AGB stars from our present M (AGB+RGB) sample, we obtain order of magnitude estimates using two approaches. 

Durrell et al. (2001) state that, at any given magnitude below the TRGB, 22\% of the stars are on the AGB as estimated from evolutionary models. We can therefore assume that 11$\pm$2 stars in our M-giant selection box are AGB stars, which leads to a C/M5+ ratio of 0.45$\pm$0.25. 
On the other hand, it is possible to turn to the AGB-fraction, $f_{\rm AGB}$, which is defined as the number ratio of luminous AGB stars above the TRGB (per magnitude) to the number of RGB stars below the tip (Armandroff et al. 1993). Accounting for the above AGB contribution below the TRGB of 22\%, we thus estimate N(RGB) as approximately 37. However, a major unknown in this reasoning is the actual value of $f_{\rm AGB}$. Observations of dSphs find a wide range of AGB fractions, e.g., 3--14\% (Lianou et al. 2010), while Martinez-Delgado \& Aparicio (1997) find 8--20\% based on CMD modeling of old populations. Here we tentatively adopt 15\% as a reasonably representative mean value, which predicts  5$\pm$1 luminous AGB stars in a sample like ours. Only accounting for these would result in a higher C/M5+ ratio of 1.0$\pm$0.7. We note, however,  that both methods assume that the number of C-stars is not affected by any of the above arguments and remains constant at five. 

To investigate the influence that incompleteness could have on our C-star counts, we determined empirical incompleteness factors using C-star tabulations
from the literature. For a number of dwarf galaxies, we determine the number ratio of luminous AGBs (above the TRGB) to the fainter C-stars below the tip. 
This ratio gives us an order of magnitude for our underestimate of the luminous AGB stars that we might miss,  {\em assuming our survey was 100\% incomplete 
above the RGB tip}. Harbeck et al. (2004) surveyed several dSphs satellites of M31 and find {\em more} C-star candidates below the TRGB. The data of Nowotny et al. (2003) for NGC 147 and NGC 185 indicate approximately a factor of two more C-stars above the TRGB compared to the fainter ones, while Battinelli \& Demers (2004a,b) find of the order of three times more luminous than fainter C-stars in the same galaxies\footnote{Battinelli et al. (2003) state that 10\% of their C-stars in the M31 disk are fainter than $I_0>20.6$, that is, the TRGB. While such an empirical incompleteness correction to our survey would yield a much higher C/M ratio of $>$1, it is interesting to note that our value found in the halo is in good agreement with that found in the disk by Battinelli et al. (2003).}. 

If we were to combine all the (extreme) effects described in this section into an empirical correction to our measurements, we could reproduce
 values for M31's C/M5+ ratio as high as $\sim$3. 
However, since the above treatment of possible incompleteness in our spectroscopic sample relies on several crude assumptions, we will retain in what follows the original value of  (0.10$\pm$0.03) determined in Sect.~4.1.
\subsection{Radial variations and velocities}
In Fig.~6 we show the radial distributions of C- and M-stars, which is clearly aggravated by the small number statistics of either component.  
In particular, we identified C-star candidates at distances of 11--32 kpc.
\begin{figure}[tb]
\begin{center}
\includegraphics[angle=0,width=1\hsize]{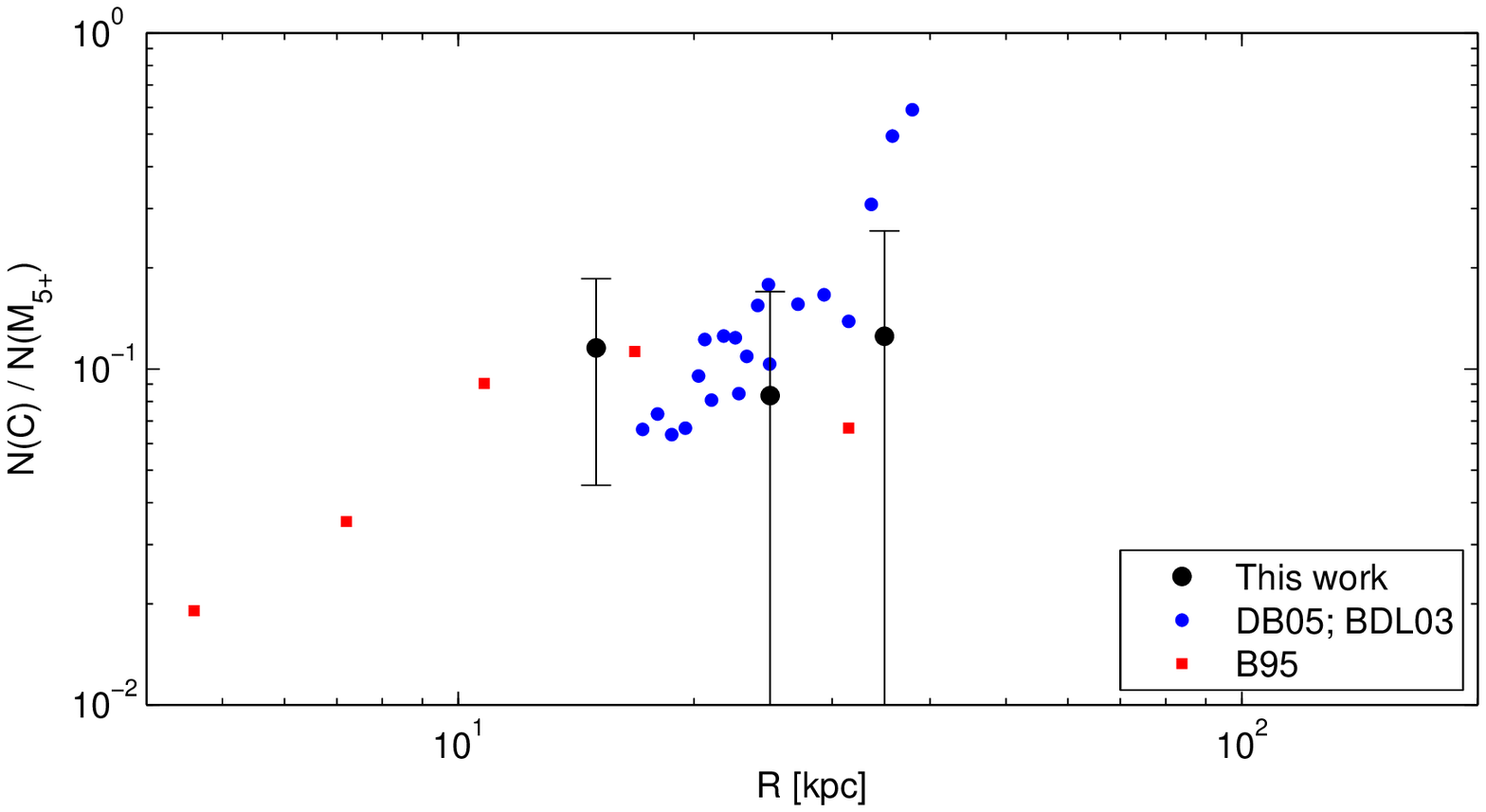}
\includegraphics[angle=0,width=1\hsize]{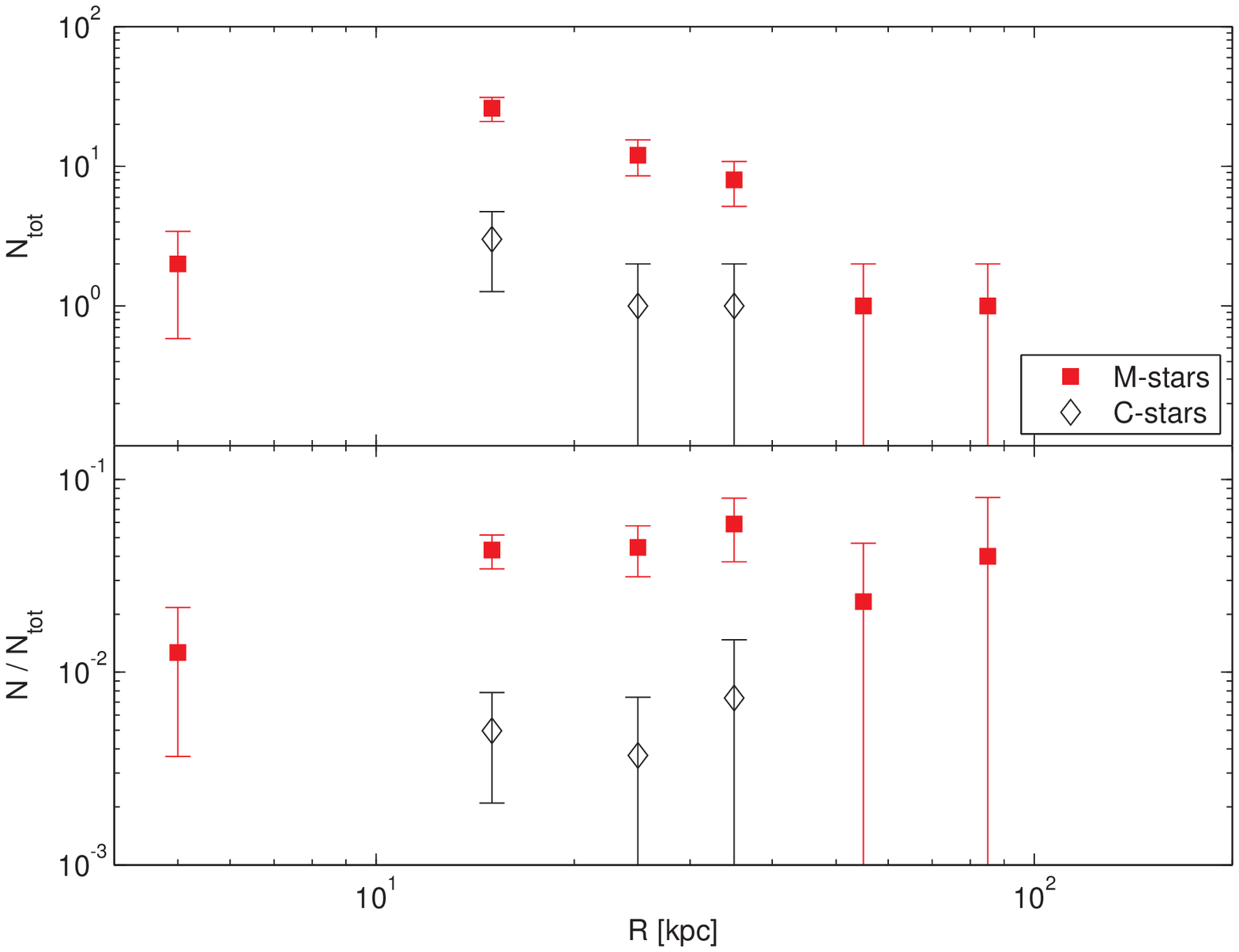}
\end{center}
\caption{Radial distribution of the stars within the empirical selection boxes. The top panel shows the overall C/M5+ ratio, while the 
middle and bottom panels shows the absolute numbers of either population and the ratio of the C- and M-stars relative to the total number of giants 
in each radial bin, respectively. Error bars are based on Poisson statistics. Also shown in the top panel are the measurements in the M31 {\em disk} by Brewer et al. (1995; B95), Battinelli et al. (2003; BDL03), and Demers \& Battinelli (2005; DB05),  although we note different selection criteria between these studies, see Sect.~4.3.3.}
\end{figure}
In those three radial bins that contain a non-zero number of carbon-stars, these correlate well with the respective M-giant populations so that the 
resulting C/M5+ ratio in each bin remains constant at $\sim$0.08--0.13. 
It is further interesting that, while the absolute numbers of M-stars in the outer fields (R$>$10 kpc) drop with radius and therefore qualitatively trace the overall, decreasing halo density profile (e.g., Ibata et al. 2007), the fraction of the M-giants relative to the entire (K- and M-) giant population is also invariant with distance, at $\sim$0.04$\pm$0.02. 

Gilbert et al. (2006), Ibata et al. (2007), and K08  detect stars that belong to the M31 halo out to large 
distances, albeit in low numbers. Their fields between 90 and 160  kpc, for instance,  were found to contain a total of 
42 red giant members (K08), 26 of which have metallicity measurements. As Fig.~6  implies, there is  only one interfering M-giant 
candidate present amongst the overall giant sample at 90 kpc, while their majority (2$\sigma$) resides within 34 kpc. 
From these low numbers, we also conclude that any possible remaining ``contamination'' of our metallicity sample with 
the few cooler, metal rich M-giants did not falsify the metallicities in none of the radial bins in K08, therefore confirming the finding of very metal poor stars and 
the strong radial metallicity gradient of K08. The presence of such a gradient and the low metallicities in the M31 halo is, however, in contrast to the 
low total number of C-stars we detect in this study (see discussion in the next section).

To investigate whether the M-giants are representative of the overall M31 halo population, we show in Fig.~7 velocity histograms of the entire 
giant sample (as solid lines) using the kinematic data of K08. These have been scaled to the total number of the M-giants, which are shown as 
shaded histograms. 
\begin{figure}[tb]
\begin{center}
\includegraphics[angle=0,width=1\hsize]{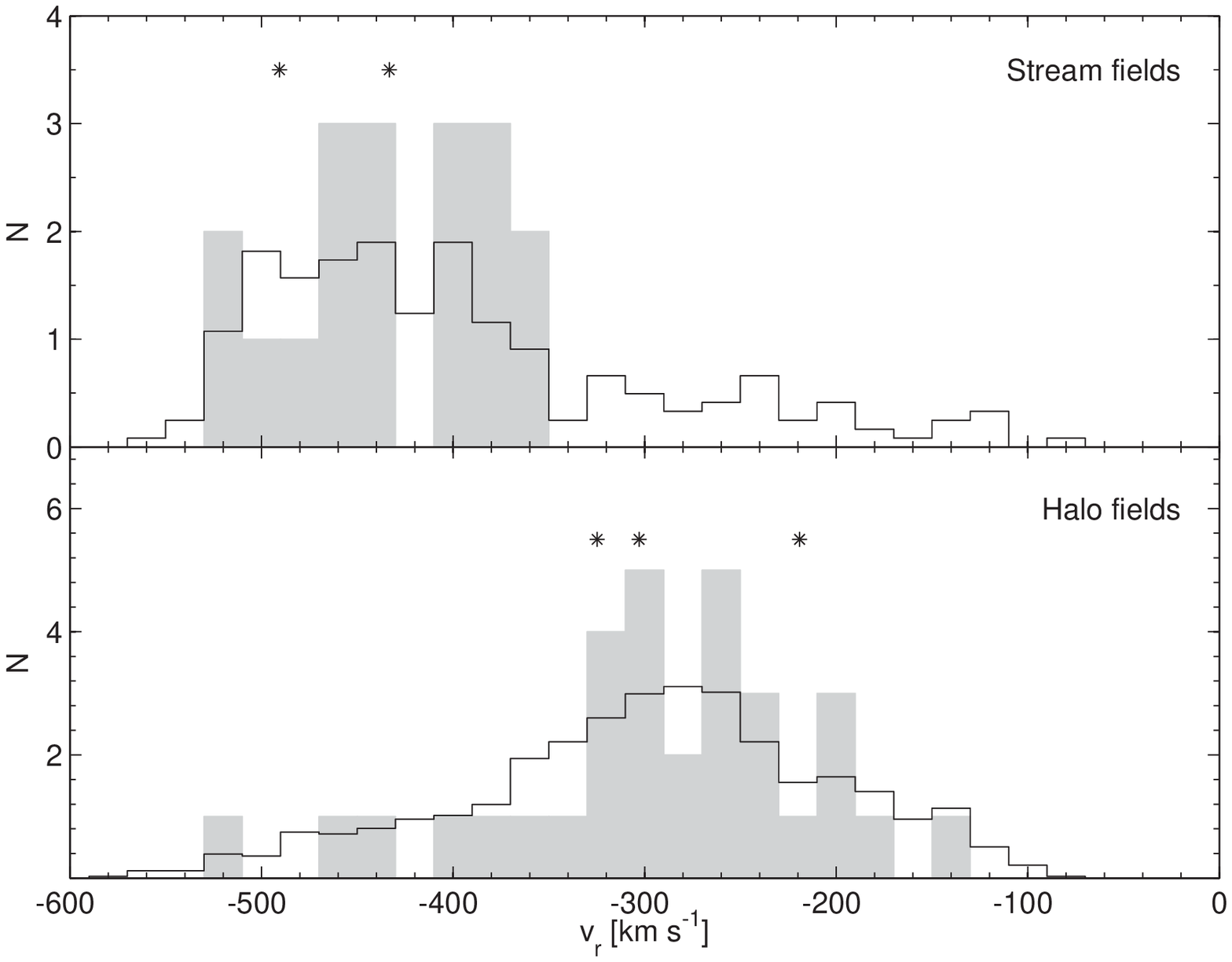}
\includegraphics[angle=0,width=1\hsize]{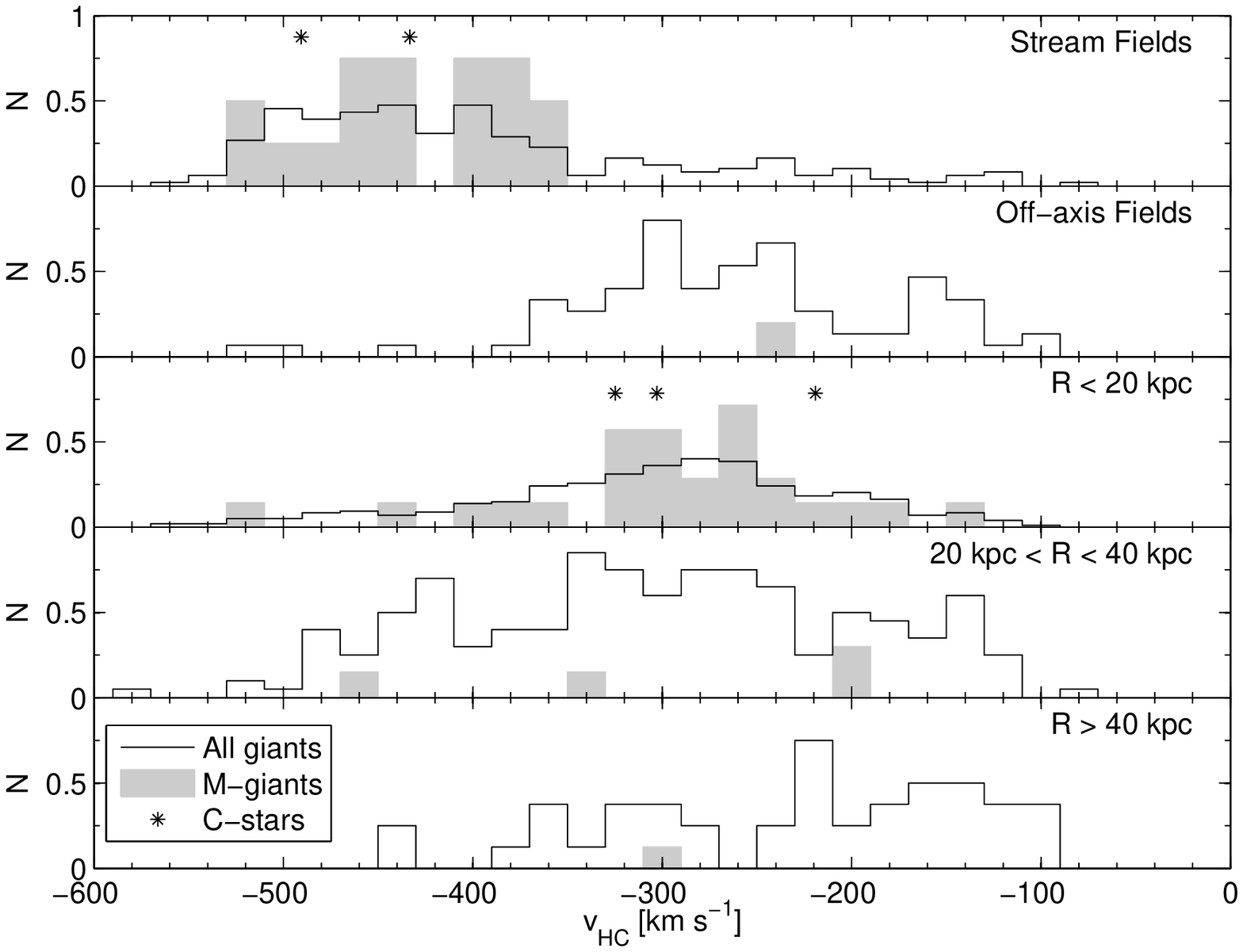}
\end{center}
\caption{Radial velocity histograms of M-giants (shaded) and all giants in our sample (solid lines), scaled to the total number of M stars (top panel). Indicated as asterisks are the velocities of the five carbon star candidates. Here, the halo fields refer to all minor- and off-axis fields at distances from 9--160 kpc, while  the stream fields are from two masks on the giant stellar stream (see also  Fig.~1 of K08). The bottom panel shows the same data, but separately for different radii. Due to the small numbers involved, these were illustratively scaled to unity.}
\end{figure}
In this plot, the ``halo fields'' contain all giant candidates from the 12 minor axis and three off-axis fields of K08, while the ``stream fields'' refer to masks H13s and a3 on top of the giant stellar stream. It is intriguing how well the M-stars trace the overall velocity distribution of the entire giant sample, both for the halo {\em and} the giant stream -- in either case, the M-star and the giant velocities are in full agreement. In the halo fields, most of the oxygen-rich stars are found at the M31 systemic velocity (at around $-$300 km\,s$^{-1}$), while the stream fields display  M-giant peaks at the stream components' velocities of $-400$ and $-520$  km\,s$^{-1}$. The same holds for the five carbon-star candidates, two of which are found in the stream fields.  
 In particular, the C/M5+ ratio in the halo (3/32; that is, 0.09$\pm$0.06) is indistinguishable from the value in the stream, at 
2/18 or 0.11$\pm$0.08. The agreement of the low value found in the Stream fields (which were preselected from only one data source) with the other halo values reassures us that this likely not an artifact of inhomogeneous selection criteria during the photometric target selections. 
On the other hand, we find a slightly higher relative number of M-giants in the stream as compared to the halo fields: while N(M)/N$_{\rm tot}$ = (8$\pm$2)\% in the 
Stream, this ratio is at (3$\pm$0.5)\% in the genuine halo fields. 
%
\section{Discussion}
We have analyzed a total of 1288 spectra of  giant stars in the M31 halo, with projected distances
ranging from 9 to 160 kpc.  Using mock narrow-band filter convolutions, we have been able to plot these giants' populations in the CN$-$TiO
vs. V$-$I diagram; from this we find that the number ratio C/M5+ is less than 0.1 in {\em all halo fields}. We detect a total of only five
C-star candidates. 
Therefore,  the outer halo exhibits the low
C/M ratio that typifies other fields that have been targeted in M31 (Brewer et al. 1995; Nowotny et al. 2001;  Batinelli \& Demers 2005). Moreover, a few of the M31 dSph satellites only show  null-detections or upper limits of C-star numbers (Groenewegen 2002; Harbeck et al. 2004). 
In contrast, disk fields closer than 30 kpc (Batinelli \& Demers 2003) host far greater absolute numbers of carbon stars  that roughly trace
the disk surface brightness. 

 If we assume the mean metallicity (and 1$\sigma$ spread) of those fields, which harbor these candidates (at $<$[Fe/H]$>$=$-1.5\pm0.41$ dex for the halo fields and 
 $-1.24\pm0.30$ dex for the Stream fields; K08), this low C/M5+ ratio in the  M31 halo fields falls $\sim 1$ order of magnitude below the well known 
 correlation of Local Group galaxies  (Fig.~8), which is defined over a broad range of morphological types by systems like the
SMC, M33, and NGC 205 (Groenewegen 2002; Mouhcine \& Lan\c con 2003; Cioni \& Habing 2003).  It is noteworthy, though, that the M31 dwarf galaxy satellites And~VII  and NGC~147 (Battinelli \& Demers 2004a; Harbeck et al. 2004) 
fall near our M31 halo and Stream fields in this plot\footnote{Nowotny et al. (2003) measure a C/M5+ ratio for NGC~147 that is higher by an order of magnitude than the value in Harbeck et al. (2004), which may be due to the different areas between these studies.}.
\begin{figure}[tb]
\begin{center}
\includegraphics[angle=0,width=1.1\hsize]{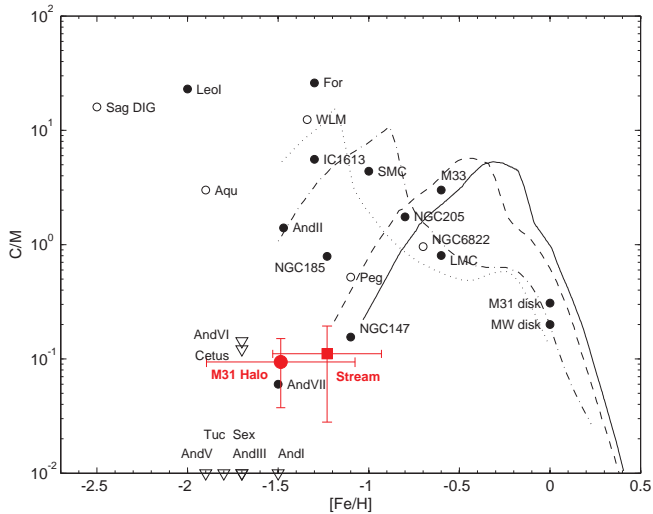}
\end{center}
\caption{C/M ratios of Local Group galaxies from the compilation of Groenewegen 2002, supplemented with data for And VII, NGC~147, NGC~185, WLM, and AndII by 
Nowotny et al. (2003), Harbeck et al. (2004),  Battinelli \& Demers (2004b), and Kerschbaum et al. (2004). Open circles indicate that only C/M0+ ratios are available in the literature, while filled symbols refer to C/M5+. Open triangles indicate C-star null-detections, or lower limits in the M-star population, in the Groenewegen, or  Harbeck et al.  data, respectively. Our measurement for the M31 halo fields is shown as a red filled circle at the mean metallicity of the fields surrounding the C-star candidates, while the solid square is the value from the two Stream fields. 
Black lines show the models of Mouhcine \& Lan\c con (2003) for different star formation histories.}
\end{figure}
NGC~147 is an M31 dwarf elliptical satellite, devoid of dust and gas (Grebel et al. 2003), that exhibits a star formation history similar to the Magellanic Clouds or the Galactic dSphs Fornax and Leo~I (Harbeck et al. 2004) in that it is massive enough to form stars over most of its life time (M$_V$=15.1). 
The M31 dSph companion And~VII (Grebel \& Guhathakurta 1999), on the other hand, is different from the remainder of the M31 dSphs in that it experienced prolonged star formation and it shows a predominant intermediate-age population of the order of 3--5 Gyr (Harbeck et al. 2004). 
These galaxies and our halo value also lie along lines of similar star formation histories, as indicated by the models by Mouhcine \& Lan\c con (2003; black lines in Fig.~8), in particular those that are characterized by a higher dominance of the younger populations (solid and long-dashed lines).
The agreement of our C/M ratio in the M31 halo with those of NGC 147 and And VII is then consistent with the interpretation that such (dSph and dE) galaxies are likely connected to some of the distinct merger features in M31;  the structures and mean stellar metallicites of these features (Ferguson et al. 2002; Harbeck et al. 2004) are further consistent with this picture.

Owing to the pre-selection criteria for the present spectroscopic sample, we cannot exclude the possibility that we have underestimated the number of 
M-type AGB stars (Sect.~4.3.3). Under realistic assumptions, however, we showed that the C/M5+ ratio is unlikely to be significantly larger than 1.0, unless we missed a substantial C-star population in our sample. 
This value is then still lower than expected from a SMC-type population and rather similar to the other M31 dwarf elliptical satellite, NGC~185. In order to reach a C/M ratio as high as  predicted by the models of Mouhcine \& Lan\c con (2003) at the given halo metallicity, M31's AGB-fraction, $f_{\rm AGB}$, would have to be remarkably low, of the order of $\la2$\%.
Moreover, only if our survey was 100\% incomplete at the bright end could we attain SMC-like C/M ratios of $\sim$3 (following our methodology in Sect.~4.3.3.). 

Prior surveys of the M31 disk, extending to 40 kpc where the halo likely dominates,
find compelling evidence that the C/M5+ fraction actually drops at the boundary of the
disk (Demers \& Battinelli 2005).  The G1 cloud of Ferguson et al. (2002), once thought to be the likely remnant
of a dwarf spheroidal satellite, exhibits a C/M5+ ratio typical of our halo fields.  
Deep HST observations (Brown et al. 2003, 2006, 2007, 2008; Richardson et al. 2009), 
find a widespread intermediate age population in the M31 halo, while Kalirai et al. (2006) and K08
both find evidence of a metal poor outer halo in M31.  Low metallicity and a strong intermediate
age population would appear to be strong factors favoring a large carbon star population.
Indeed, this appears to be the case for the Galactic dwarf spheroidals (e.g., Azzopardi et al. 1985,1986; Groenewegen 1999; Demers \& Battinelli 2002).  
However, the Galactic halo does not appear to be comprised of the debris of
such systems with major C-star populations.  Even if the parent systems disrupted, the carbon-rich stars should have survived in the halo field (see Ibata et al. 2001b for the Sgr dwarf).  

We are left to conclude that two factors may decrease the expected numbers of carbon stars.  First, the age distribution
in the outer halo may be tilted toward older stars.   However, in the G1 clump (at a projected distance of 34 kpc), there is evidence that the field
has a strong intermediate-age component (Rich et al. 2004; Faria et al. 2007) and also an HST field at 35 kpc was shown to contain $\sim$1/3 of stars younger than 10 Gyr (Brown et al. 
2008).   
The second explanation would argue that the halo is too metal rich to support a substantial population of C-stars.  
However, both Kalirai et al. (2006) and K08  
agree that [Fe/H]$\sim -1.5$ and lower at large distances, a value that is low enough to support carbon star production; moreover, there is a dispersion in
metallicity with a metal rich tail high enough to spawn M giants.
Considering that the origin of this metal rich tail must arise in mergers, the absence of a strong carbon star population in the outer halo must be considered to be noteworthy.  
Therefore, the low C/M ratio in the M31 field is inconsistent with a presumed disk-like or dwarf galaxy origin (with properties like those indicated in Fig.~8) of those halo fields over a broad radial extent. 
Although the total numbers of stars are low in the halo, the C/M {\em ratio} should be preserved.  

The cool M-giants with detectable TiO, on the other hand, are found throughout the halo, further strengthening the case for a widespread 
population with [Fe/H]$>-1$, even for the most distant giants.  
Further, stars with measureable TiO follow the same kinematics as is found for all giants with measurable radial velocities. 
Both regarding all M stars and with respect to the C/M ratio (as measured by our very small numbers) the 
giant stream appears to harbor no special stellar population, in contrast to what has been found for the Sgr
dwarf spheroidal galaxy (Ibata et al. 2001b), which does have a concentration of carbon stars.

With the explosive discovery of numerous streams and stellar systems in the halo, the picture of the halo population comprised nearly entirely of stellar subsystems has become the canon.  Yet well constructed surveys of our Milky Way halo (e.g. Mauron et al. 2008) find only a handful of luminous carbon stars.  If indeed halos are dominated by debris, as cannot ruled out from a dynamical standpoint, they are not similar in population to the SMC, NGC 6822, or M33.  If we appeal to the disruption of a disk-like galaxy to have created the giant stream (e.g., bearing on the model of Fardal et al. 2008), that progenitor was especially devoid of carbon stars and more closely resembled NGC 147 rather than M33.  
In the broader picture, the late-type stellar content of the M31 halo must be judged to differ from that of the dwarf galaxies that one
 might propose to have contributed its stars.
\acknowledgments
We would like to thank M.-R. Cioni for helpful discussions and an anonymous referee for a constructive report. 
AK acknowledges support by an STFC postdoctoral fellowship. 
Support was provided by NSF (AST-0307931, AST-0709479),  
HST (GO-10265, 10816), and by R.M. Rich. 
Some of the data presented herein were 
obtained using the W. M. Keck Observatory, which is operated 
as a scientific partnership among Caltech, the University of 
California, and NASA. The Observatory was made possible 
by the generous financial support of the W. M. Keck 
Foundation. 
This research used the facilities of the
Canadian Astronomy Data Centre operated by
the National Research Council of Canada
with the support of the Canadian Space Agency.
\begin{appendix}
Here we show the TCDs separately for different observing runs and different targeted fields in order to test for photometric homogeneity, as discussed in Sect.~4.3.2.
\begin{figure*}[htb]
\begin{center}
\includegraphics[angle=0,width=0.3\hsize]{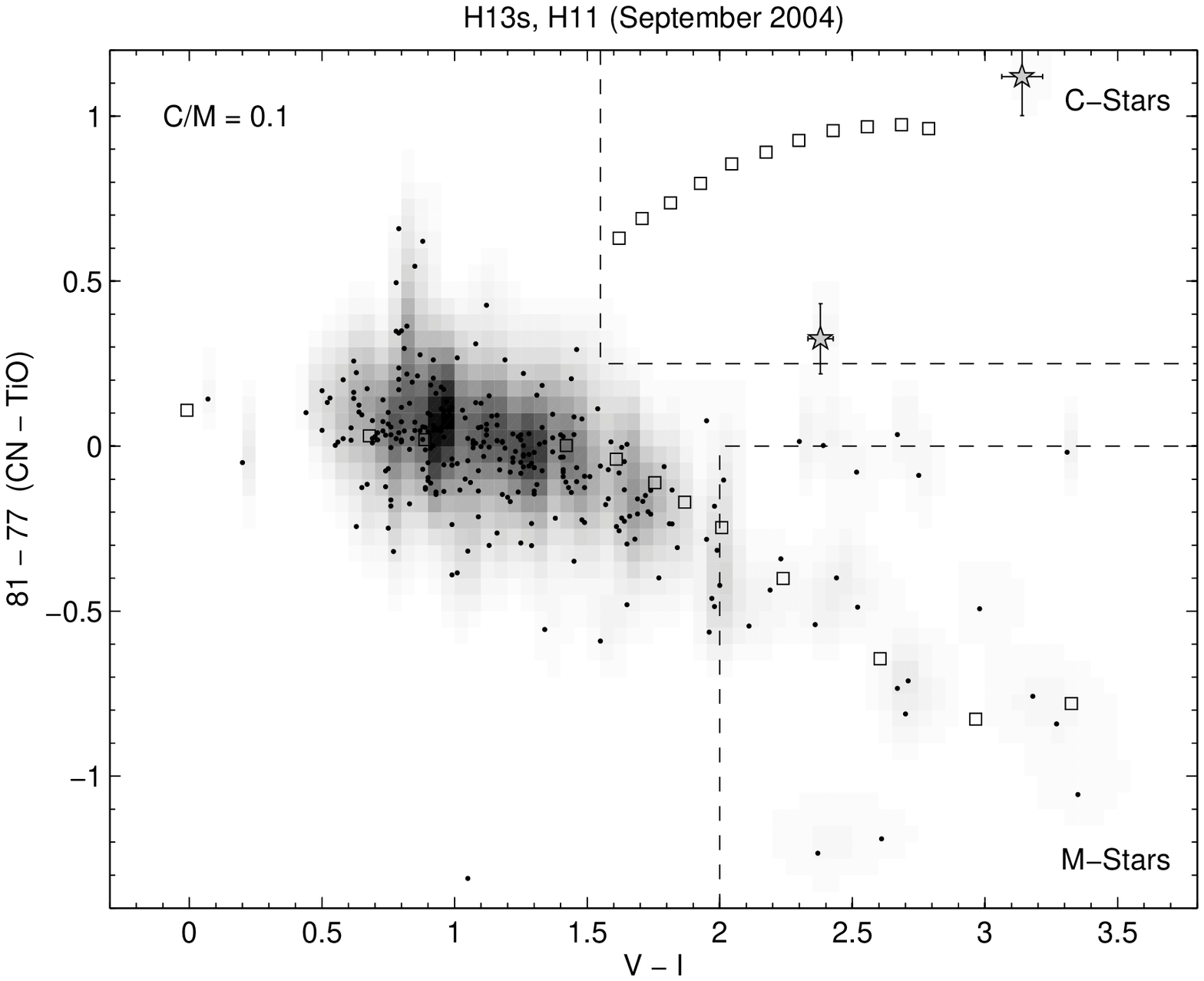}
\includegraphics[angle=0,width=0.3\hsize]{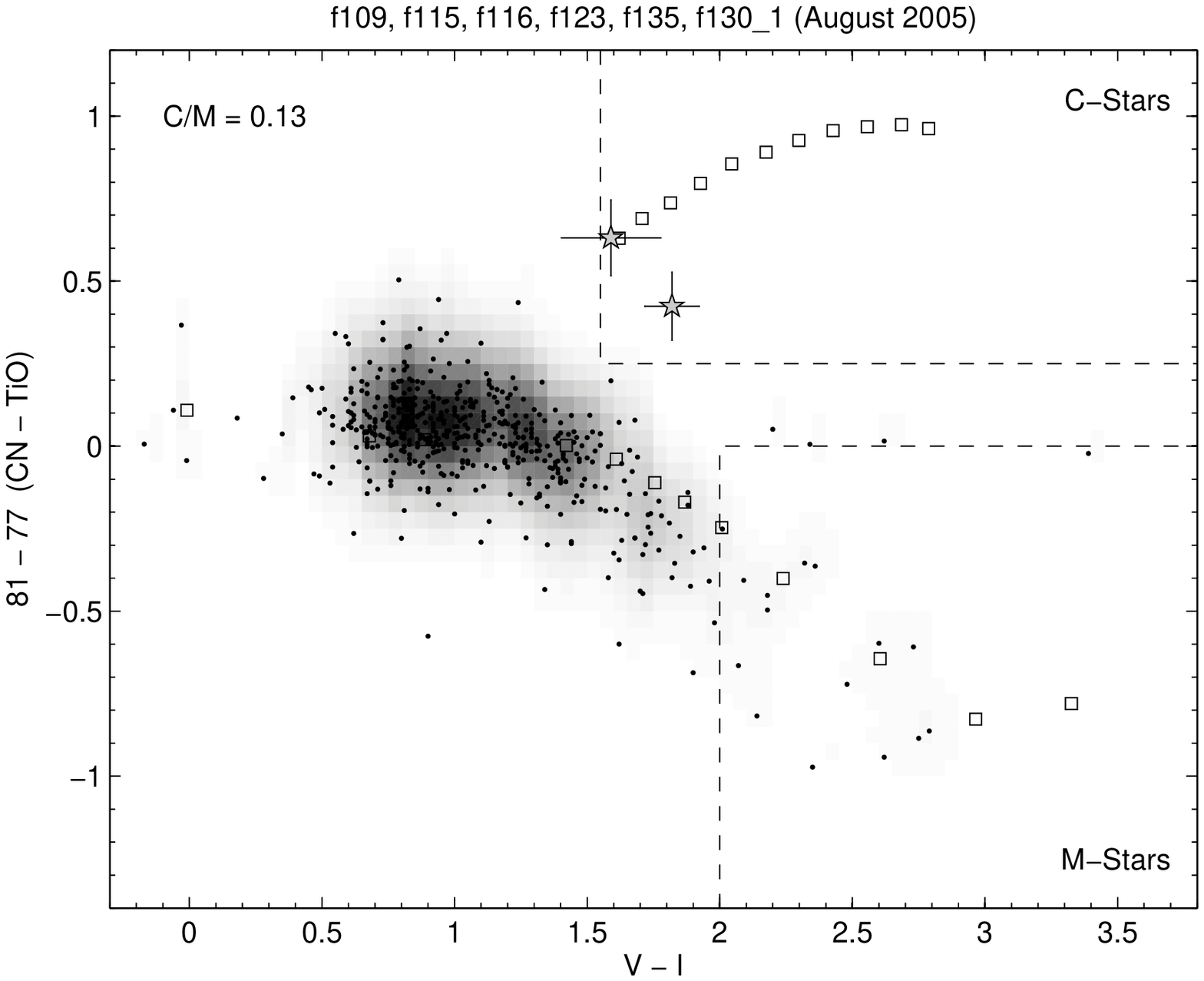}
\includegraphics[angle=0,width=0.3\hsize]{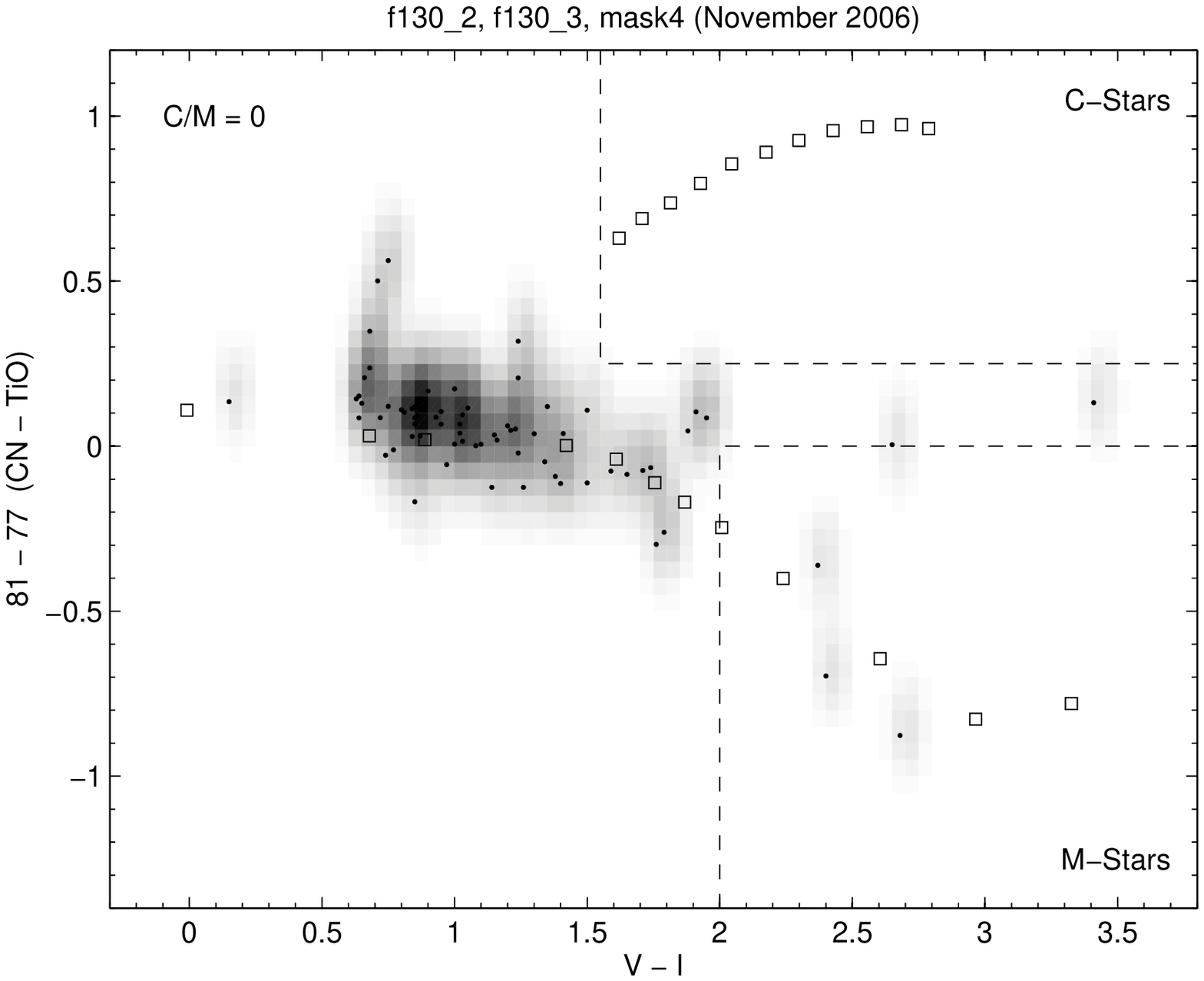}\\
\includegraphics[angle=0,width=0.3\hsize]{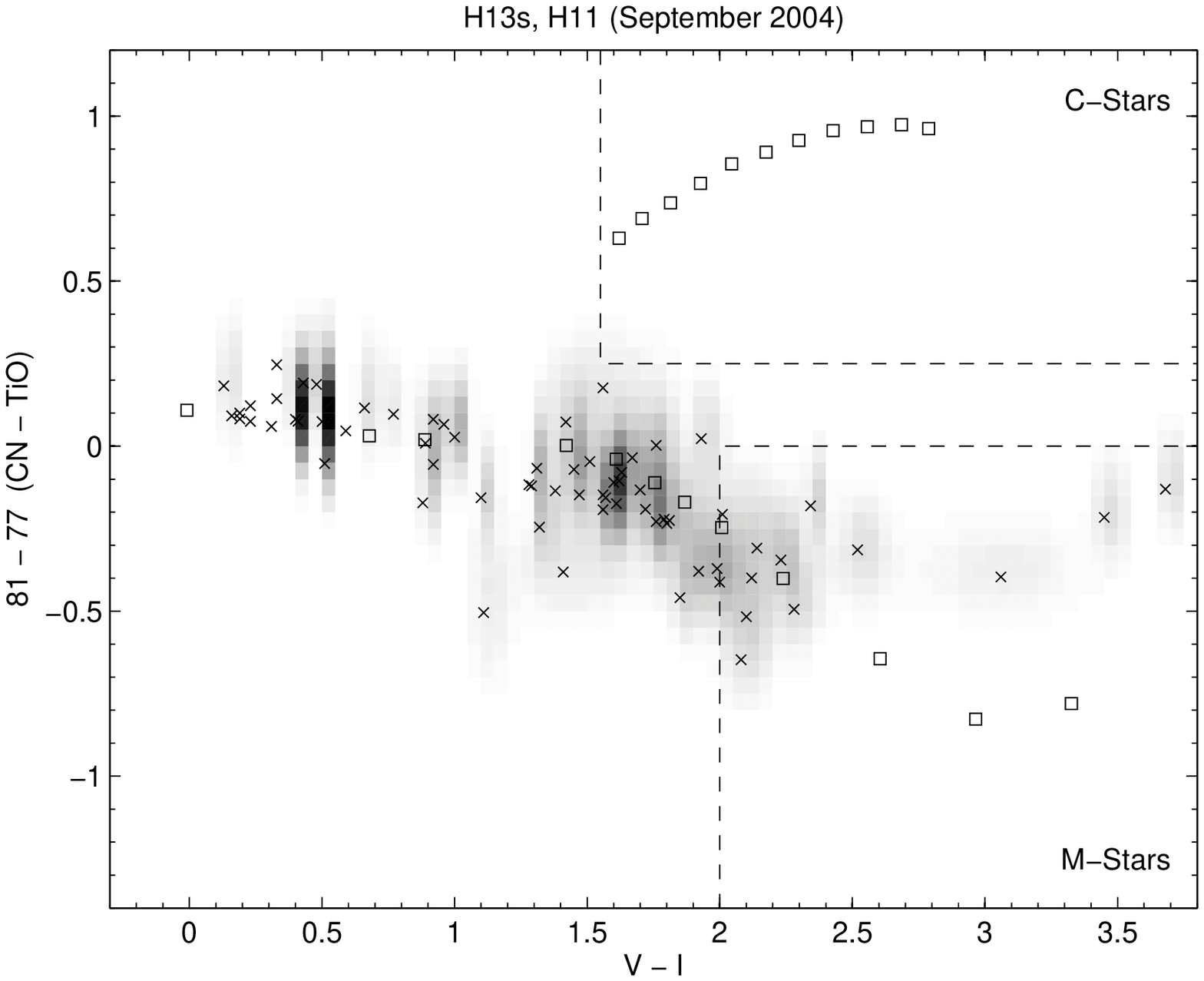}
\includegraphics[angle=0,width=0.3\hsize]{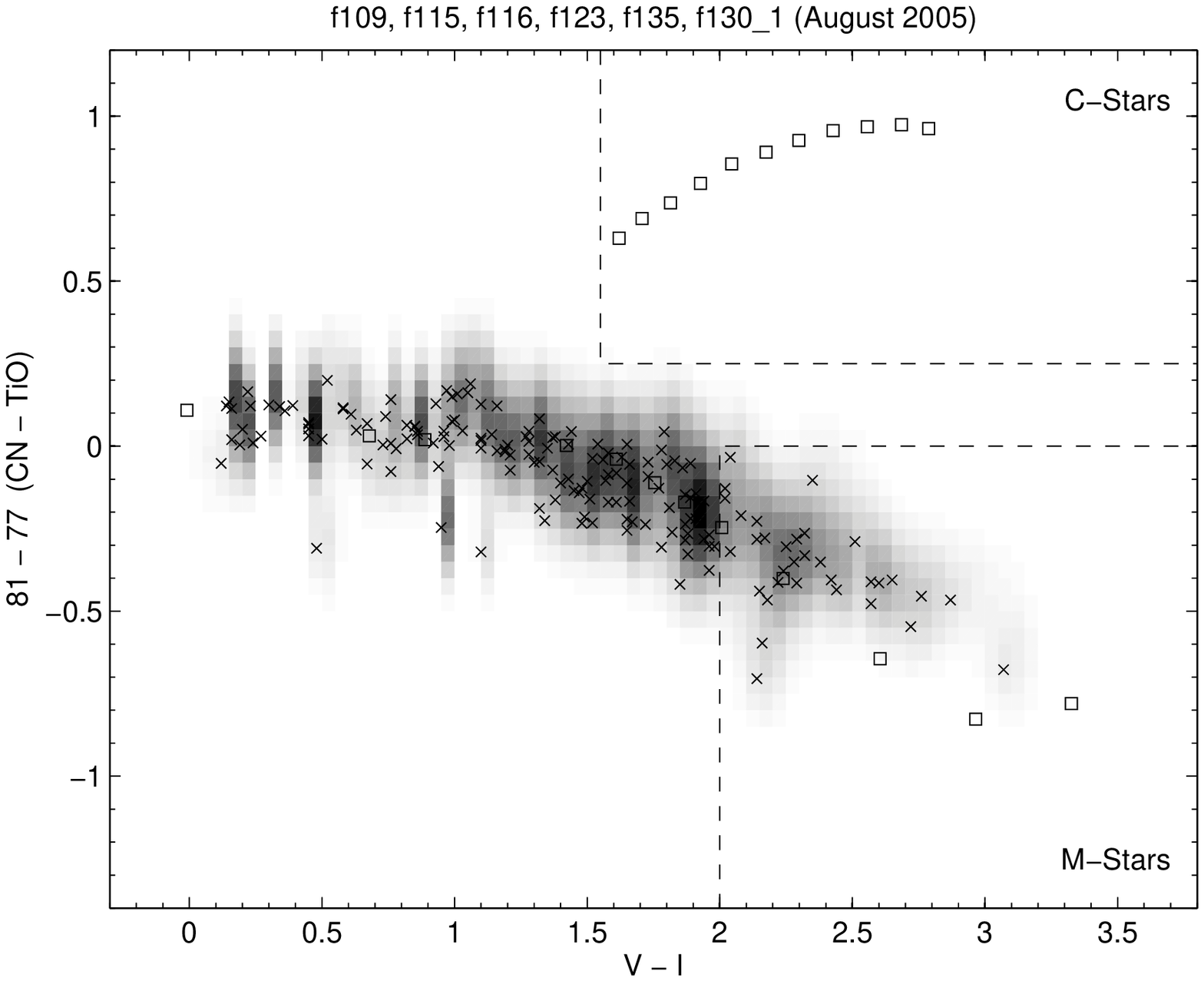}
\includegraphics[angle=0,width=0.3\hsize]{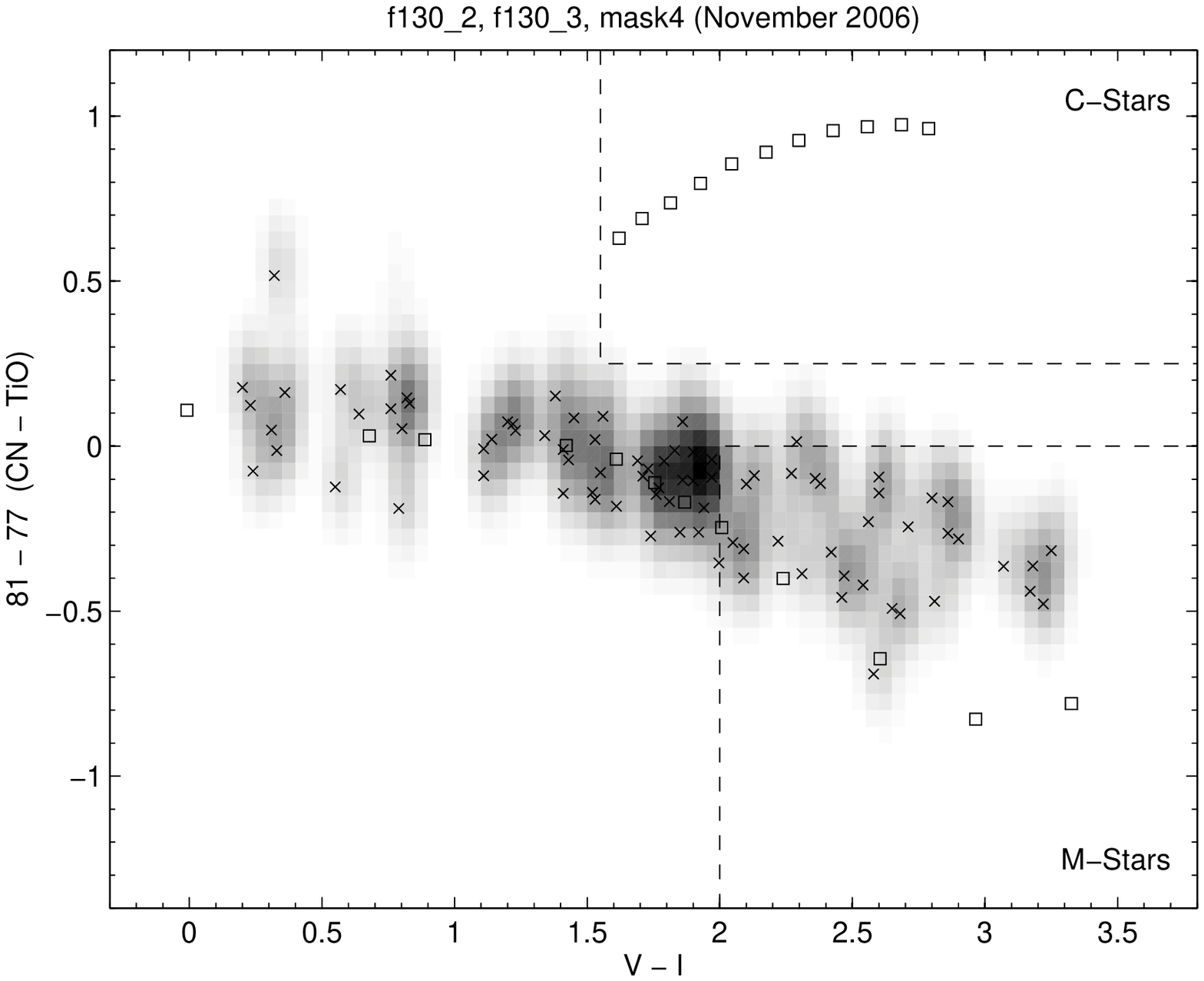}
\end{center}
\caption{Same as Fig.~4, but separately for each of the three observing runs that covered the inner halo fields (R$<$20 kpc) and
that has homogenous photometry from the MegaCam survey. Shown at the top are giants and the bottom panels are for foreground dwarfs. 
To aid the comparison of the diagrams, we also show in grayscale the individual number distributions.}
\end{figure*}
\begin{figure*}[htb]
\begin{center}
\includegraphics[angle=0,width=0.4\hsize]{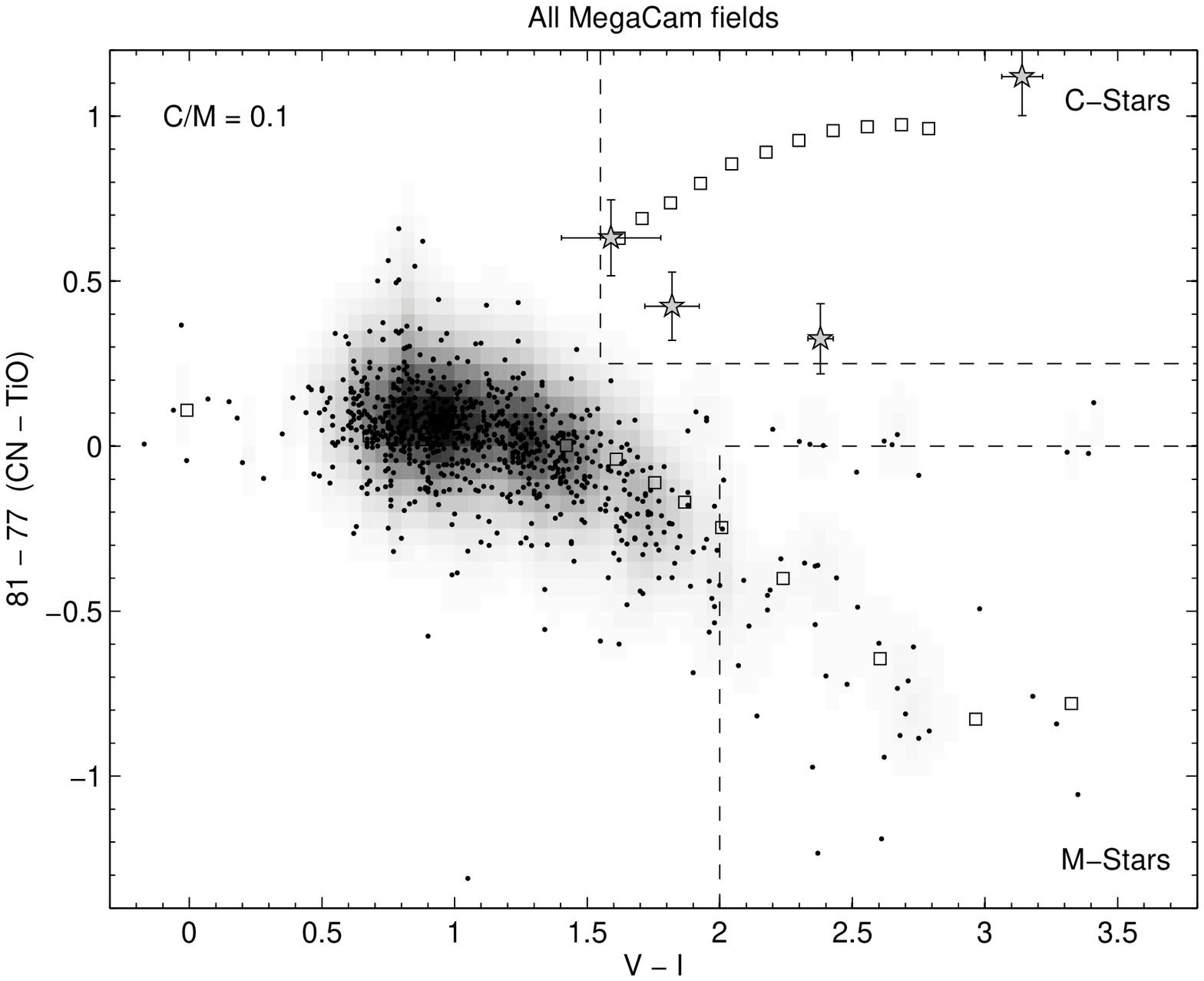}
\includegraphics[angle=0,width=0.4\hsize]{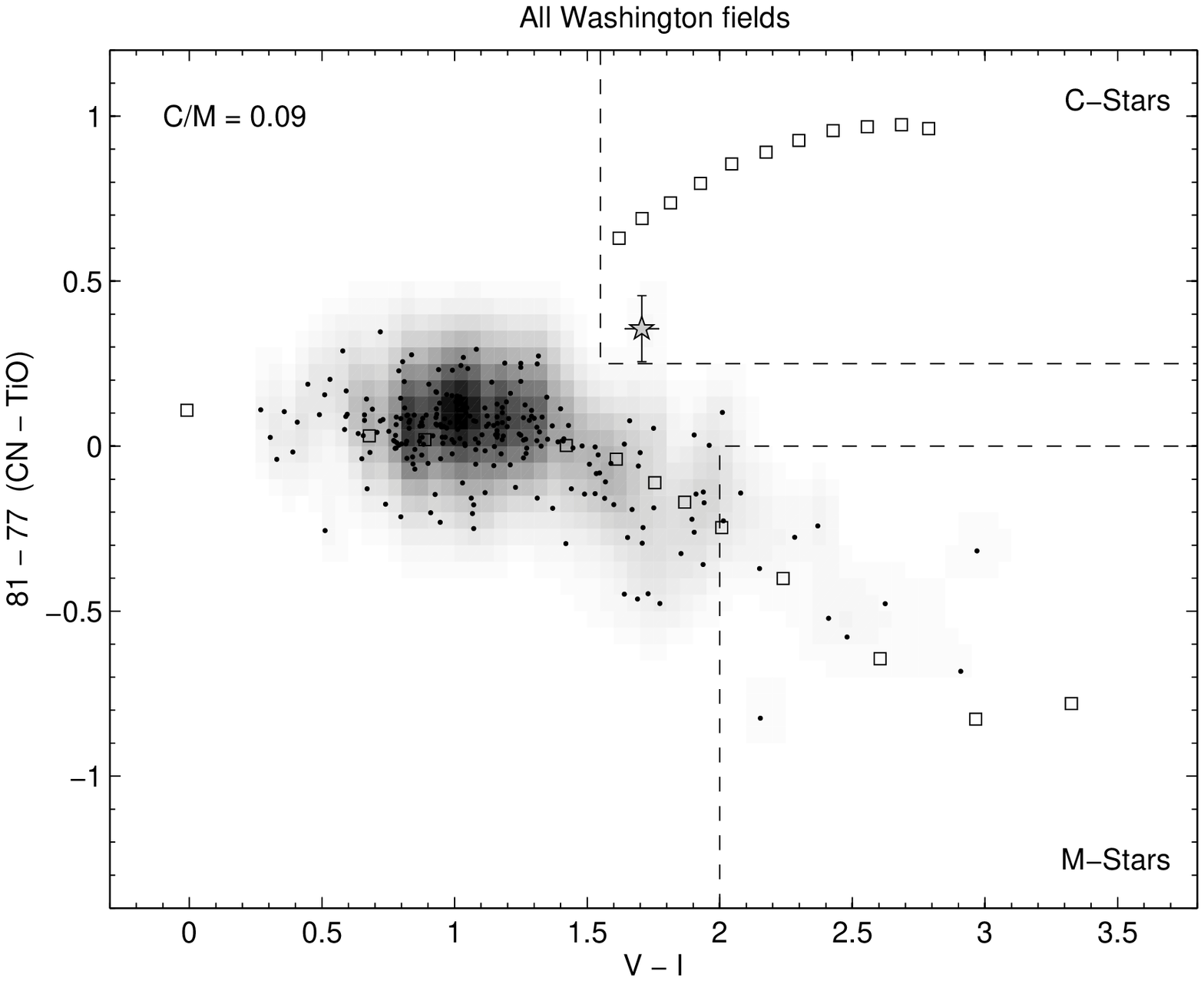}
\includegraphics[angle=0,width=0.4\hsize]{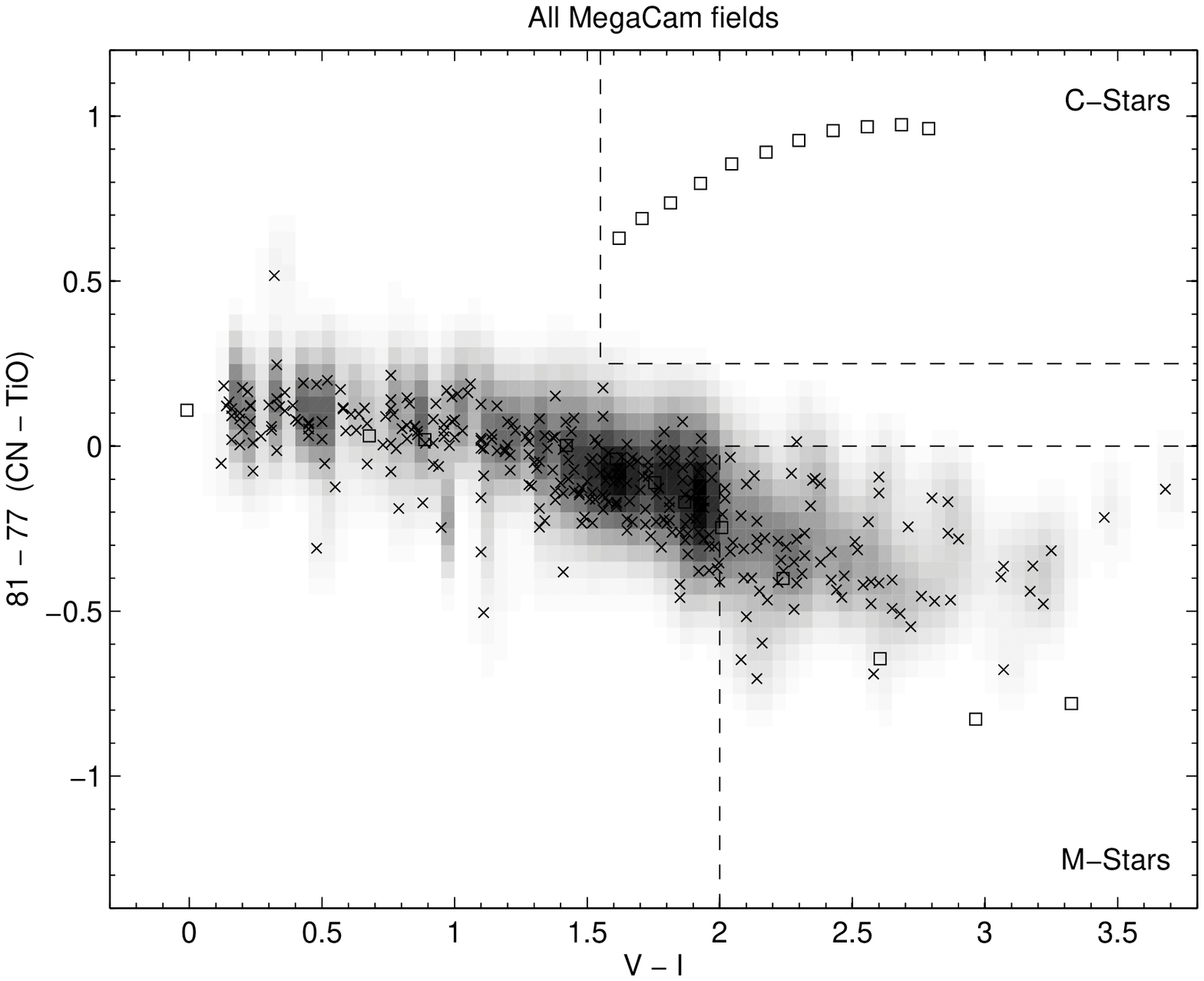}
\includegraphics[angle=0,width=0.4\hsize]{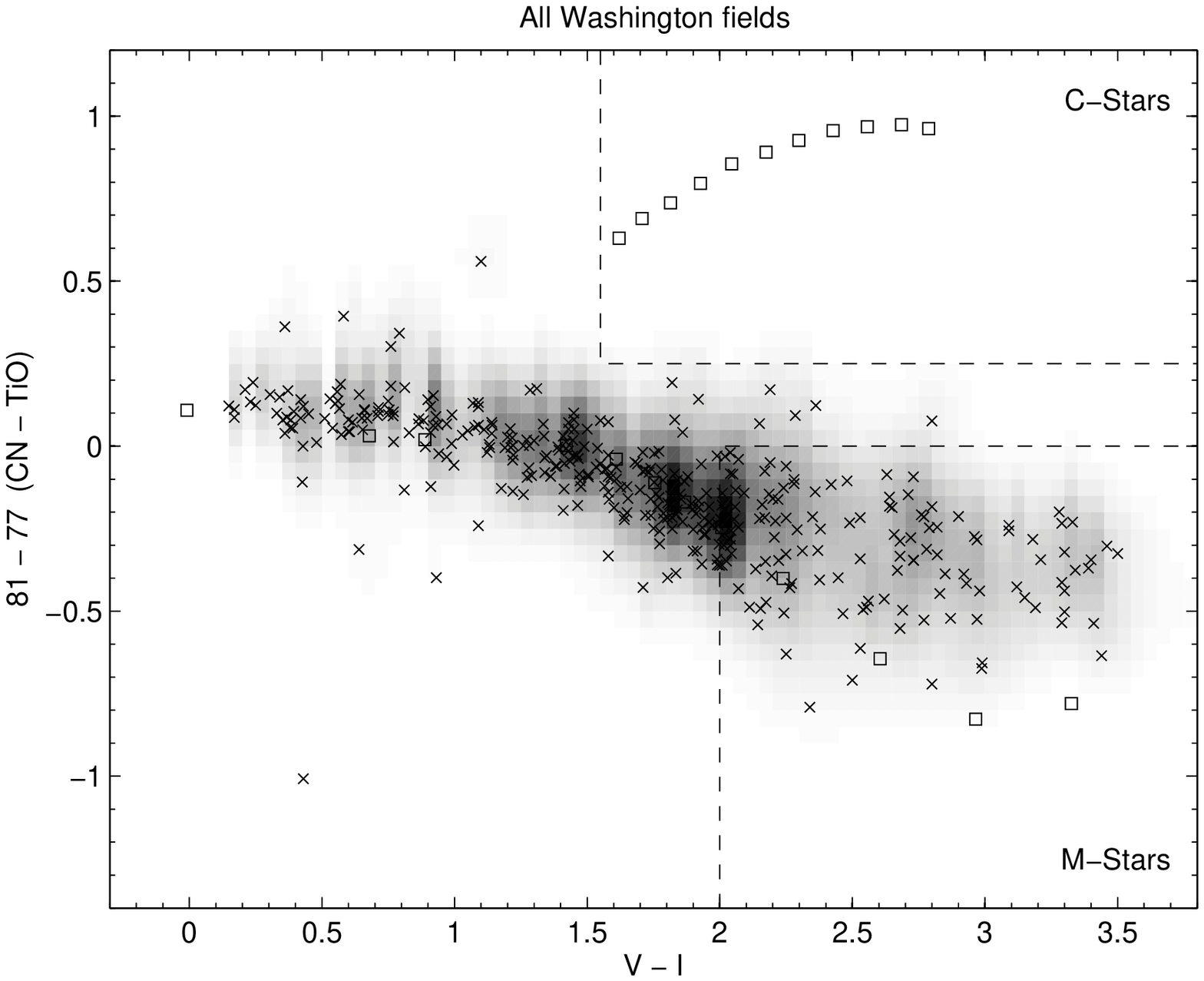}
\end{center}
\caption{Same as Fig.~9, but separately for all fields based on the pre-selection from MegaCam photometry (left panels) versus all outer fields 
in which the spectroscopic targets have been selected from Washington photometry (right). }
\end{figure*}
\end{appendix}

\begin{thebibliography}{}
%
\bibitem[Armandroff et al.(1993)]{1993AJ....106..986A} Armandroff, T.~E.,  Da Costa, G.~S., Caldwell, N., \& Seitzer, P.\ 1993, \aj, 106, 986 
%
\bibitem[Azzopardi et al.(1986)]{1986A&A...161..232A} Azzopardi, M., Lequeux, J., \& Westerlund, B.~E.\ 1986, \aap, 161, 232 
%
\bibitem[Azzopardi et al.(1985)]{1985A&A...144..388A} Azzopardi, M., Lequeux, J., \& Westerlund, B.~E.\ 1985, \aap, 144, 388 
%
\bibitem[Battinelli et al.(2003)]{2003AJ....125.1298B} Battinelli, P., Demers, S., \& Letarte, B.\ 2003, \aj, 125, 1298 
%
\bibitem[Battinelli \& Demers(2004)]{2004A&A...418...33B} Battinelli, P., \& Demers, S.\ 2004a, \aap, 418, 33 
%
\bibitem[Battinelli \& Demers(2004)]{2004A&A...416..111B} Battinelli, P., \& Demers, S.\ 2004b, \aap, 416, 111 
%
\bibitem[Battinelli \& Demers(2005)]{2005A&A...430..905B} Battinelli, P., \& Demers, S.\ 2005, \aap, 430, 905 
%
\bibitem[Belokurov et al.(2007)]{2007ApJ...658..337B} Belokurov, V., et al.\ 2007a, \apj, 658, 337 
%
\bibitem[Belokurov et al.(2007)]{2007ApJ...654..897B} Belokurov, V., et al.\ 2007b, \apj, 654, 897 
%
\bibitem[Blanco et al.(1984)]{1984AJ.....89..636B} Blanco, V.~M., McCarthy,  M.~F., \& Blanco, B.~M.\ 1984, \aj, 89, 636 
%
\bibitem[Brewer et al.(1995)]{1995AJ....109.2480B} Brewer, J.~P., Richer, H.~B., \& Crabtree, D.~R.\ 
1995, \aj, 109, 2480 
%
\bibitem[Brewer et al.(1996)]{1996AJ....112..491B} Brewer, J.~P., Richer,  H.~B., \& Crabtree, D.~R.\ 
1996, \aj, 112, 491 
%
\bibitem[Brown et al.(2003)]{2003ApJ...592L..17B} Brown, T.~M., Ferguson, H.~C., Smith, E., Kimble, R.~A., Sweigart, A.~V., Renzini, A., Rich, R.~M., \& VandenBerg, D.~A.\ 2003, \apjl, 592, L17 
%
\bibitem[Brown et al.(2006)]{2006ApJ...652..323B} Brown, T.~M., Smith, E., Ferguson, H.~C., Rich, R.~M., Guhathakurta, P., Renzini, A., Sweigart, A.~V., \& Kimble, R.~A.\ 2006, \apj, 652, 323 
%
\bibitem[Brown et al.(2007)]{2007ApJ...658L..95B} Brown, T.~M., et al.\ 
2007, \apjl, 658, L95 
%
\bibitem[Brown et al.(2008)]{2008ApJ...685L.121B} Brown, T.~M., et al.\ 
2008, \apjl, 685, L121 
%
\bibitem[Chou et al.(2007)]{2007ApJ...670..346C} Chou, M.-Y., et al.\ 2007, \apj, 670, 346 
%
\bibitem[Cioni \& Habing(2003)]{2003A&A...402..133C} Cioni, M.-R.~L., \& Habing, H.~J.\ 2003, \aap, 402, 133 
%
\bibitem[Cioni et al.(2008)]{2008A&A...487..131C} Cioni, M.-R.~L., et al.\ 2008, \aap, 487, 131 
%
\bibitem[Cook et al.(1986)]{1986ApJ...305..634C} Cook, K.~H., Aaronson, M.,  \& Norris, J.\ 1986, \apj, 305, 634 
%
\bibitem[Cook \& Aaronson(1989)]{1989AJ.....97..923C} Cook, K.~H., \& Aaronson, M.\ 1989, \aj, 97, 923 
%
\bibitem[Demers \& Battinelli(2002)]{2002AJ....123..238D} Demers, S., \& Battinelli, P.\ 2002, \aj, 123, 238 
%
\bibitem[Demers \& Battinelli(2005)]{2005A&A...436...91D} Demers, S., \& Battinelli, P.\ 2005, \aap, 436, 91 
%
\bibitem[Durrell et al.(2001)]{2001AJ....121.2557D} Durrell, P.~R., Harris, W.~E., \& Pritchet, C.~J.\ 2001, \aj, 121, 2557 
%
\bibitem[Fardal et al.(2006)]{2006MNRAS.366.1012F} Fardal, M.~A., Babul, A., Geehan, J.~J., \& Guhathakurta, P.\ 2006, \mnras, 366, 1012 
%
\bibitem[Fardal et al.(2008)]{2008ApJ...682L..33F} Fardal, M.~A., Babul, A., Guhathakurta, P., Gilbert, K.~M., \& Dodge, C.\ 2008, \apjl, 682, L33 
%
\bibitem[Faria et al.(2007)]{2007AJ....133.1275F} Faria, D., Johnson,  R.~A., Ferguson, A.~M.~N., Irwin, M.~J., Ibata, R.~A., Johnston, K.~V.,  Lewis, G.~F., \& Tanvir, N.~R.\ 2007, \aj, 133, 1275 
%
\bibitem[Fellhauer et al.(2006)]{2006ApJ...651..167F} Fellhauer, M., et  al.\ 2006, \apj, 651, 167 
%
\bibitem[Ferguson et al.(2002)]{2002AJ....124.1452F} Ferguson, A.~M.~N., Irwin, M.~J., Ibata, R.~A., Lewis, G.~F., \& Tanvir, N.~R.\ 2002, \aj, 124, 1452 
%
\bibitem[Fiorucci \& Munari(2003)]{2003A&A...401..781F} Fiorucci, M., \& Munari, U.\ 2003, \aap, 401, 781 
%
\bibitem[Gilbert et al.(2006)]{2006ApJ...652.1188G} Gilbert, K.~M., et al.\  2006, \apj, 652, 1188 
%
\bibitem[Grebel \& Guhathakurta(1999)]{1999ApJ...511L.101G} Grebel, E.~K., \& Guhathakurta, P.\ 1999, \apjl, 511, L101 
%
\bibitem[Grebel et al.(2003)]{2003AJ....125.1926G} Grebel, E.~K., Gallagher, J.~S., III, \& Harbeck, D.\ 2003, \aj, 125, 1926 
%
\bibitem[Groenewegen(2002)]{2002astro.ph..8449G} Groenewegen, M.~A.~T.\  2002, in Chemical Evolution of Dwarf Galaxies, Ringberg Workshop (arXiv:astro-ph/0208449) 
%
\bibitem[Gwyn(2008)]{2008PASP..120..212G} Gwyn, S.~D.~J.\ 2008, \pasp, 120, 212 
%
\bibitem[Harbeck et al.(2004)]{2004AJ....127.2711H} Harbeck, D., Gallagher, J.~S., III, \& Grebel, E.~K.\ 
2004, \aj, 127, 2711 
%
\bibitem[Ibata et al.(2001)]{2001Natur.412...49I} Ibata, R., Irwin, M.,  Lewis, G., Ferguson, A.~M.~N., \& Tanvir, N.\ 2001a, \nat, 412, 49 
%
\bibitem[Ibata et al.(2001)]{2001ApJ...547L.133I} Ibata, R., Irwin, M.,  Lewis, G.~F., \& Stolte, A.\ 2001b, \apjl, 547, L133 
%
\bibitem[Ibata et al.(2007)]{2007ApJ...671.1591I} Ibata, R., Martin, N.~F., Irwin, M., Chapman, S., Ferguson, A.~M.~N., Lewis, G.~F., \& McConnachie, A.~W.\ 2007, \apj, 671, 1591 
%
\bibitem[Iben  \& Renzini(1983)]{1983ARA&A..21..271I} Iben, I., Jr., \& Renzini, A.\ 1983, \araa, 21, 271 
%
\bibitem[Kalirai et al.(2006a)]{2006ApJ...648..389K} Kalirai, J.~S., et al.\ 
2006, \apj, 648, 389 	
%
\bibitem[Kerschbaum et  al.(2004)]{2004A&A...427..613K} Kerschbaum, F., Nowotny, W., Olofsson, H., \& Schwarz, H.~E.\ 2004, \aap, 427, 613 
%
\bibitem[Koch et al.(2008)]{2008ApJ...689..958K} Koch, A., et al.\ 2008, 
\apj, 689, 958 (K08)
%
\bibitem[Law et al.(2005)]{2005ApJ...619..807L} Law, D.~R., Johnston, K.~V., \& Majewski, S.~R.\ 2005, \apj, 619, 807 
%
\bibitem[Letarte et al.(2002)]{2002AJ....123..832L} Letarte, B., Demers,  S., Battinelli, P., \& Kunkel, W.~E.\ 2002, \aj, 123, 832 
%
\bibitem[Lianou et al.(2010)]{2010arXiv1003.0861L} Lianou, S., Grebel, E.K., \& Koch, A. 2010, A\&A, in press (arXiv:1003.0861)
%
\bibitem[Majewski et al.(2000)]{2000AJ....119..760M} Majewski, S.~R., 
Ostheimer, J.~C., Patterson, R.~J., Kunkel, W.~E., Johnston, K.~V., \& 
Geisler, D.\ 2000, \aj, 119, 760 
%
\bibitem[Marigo et al.(2008)]{2008A&A...482..883M} Marigo, P., Girardi, L., Bressan, A., Groenewegen, M.~A.~T., Silva, L., \& Granato, G.~L.\ 2008, \aap, 482, 883 
%
\bibitem[Martinez-Delgado 
\& Aparicio(1997)]{1997ApJ...480L.107M} Martinez-Delgado, D., \& Aparicio, A.\ 1997, \apjl, 480, L107 
%
\bibitem[Mauron(2008)]{2008A&A...482..151M} Mauron, N.\ 2008, \aap, 482, 151 
%
\bibitem[Mori \& Rich(2008)]{2008ApJ...674L..77M} Mori, M., \& Rich, R.~M.\ 2008, \apjl, 674, L77 
%
\bibitem[Moro \& Munari(2000)]{2000A&AS..147..361M} Moro, D., \& Munari, U.\ 2000, \aaps, 147, 361 
%
\bibitem[Mouhcine \& Lan{\c c}on(2003)]{2003MNRAS.338..572M} Mouhcine, M., \& Lan{\c c}on, A.\ 2003, \mnras, 338, 572 
%
\bibitem[Mould  \& Aaronson(1979)]{1979ApJ...232..421M} Mould, J., \& Aaronson, M.\ 1979, \apj, 232, 421 
%
\bibitem[Nowotny et al.(2001)]{2001A&A...367..557N} Nowotny, W., Kerschbaum, F., Schwarz, H.~E., \& Olofsson, H.\ 2001, \aap, 367, 557 
%
\bibitem[Nowotny et  al.(2003)]{2003A&A...403...93N} Nowotny, W., Kerschbaum, F., Olofsson, H., \& Schwarz, H.~E.\ 2003, \aap, 403, 93 
%
\bibitem[Ostheimer(2003)]{2003PhDT.........4O} Ostheimer, J.~C.~J.\ 2003, 
Ph.D.~Thesis,  University of Virgina
%
\bibitem[Palma et al.(2003)]{2003AJ....125.1352P} Palma, C., Majewski, S.~R., Siegel, M.~H., Patterson, 
R.~J., Ostheimer, J.~C., \& Link, R.\ 2003, \aj, 125, 1352 
%
\bibitem[Rich et al.(2004)]{2004AJ....127.2139R} Rich, R.~M., Reitzel, D.~B., Guhathakurta, P., Gebhardt, K., \& Ho, L.~C.\ 2004, \aj, 127, 2139 
%
\bibitem[Richardson et al.(2008)]{2008AJ....135.1998R} Richardson, J.~C., et al.\ 2008, \aj, 135, 1998 
%
\bibitem[Richardson et al.(2009)]{2009MNRAS.396.1842R} Richardson, J.~C., et al.\ 2009, \mnras, 396, 1842 
%
\bibitem[Searle \& Zinn(1978)]{1978ApJ...225..357S} Searle, L., \& Zinn, R.\ 1978, \apj, 225, 357 
%
\bibitem[Stanek \& Garnavich(1998)]{1998ApJ...503L.131S} Stanek, K.~Z., \& 
Garnavich, P.~M.\ 1998, \apjl, 503, L131 
%
\bibitem[Totten et al.(2000)]{2000MNRAS.314..630T} Totten, E.~J., Irwin, M.~J., \& Whitelock, P.~A.\ 
2000, \mnras, 314, 630 
%
\end{thebibliography}
\end{document}